\PassOptionsToPackage{draft}{hypref}
\PassOptionsToPackage{breaklinks=true}{hypref}
\documentclass[a4paper,fleqn,usenatbib]{mnras}
\PassOptionsToPackage{draft}{hyperref}
\usepackage{graphicx}
\usepackage{amsmath}

\usepackage{supertabular}

\DeclareMathOperator\erf{erf}

\makeatletter
\@fleqnfalse
\@mathmargin\@centering
\makeatother

\author[D. J. Prole et al.]{D. J. Prole,$^{1}$\thanks{ProleD@Cardiff.ac.uk}
									   J. I. Davies,$^{1}$
									   O. C. Keenan$^{1}$
									   and L. J. M. Davies$^{2}$
									   \\
									   $^{1}$School of Physics and Astronomy, Cardiff University, The Parade, Cardiff, CF24
									   3AA, UK \\
									   $^{2}$ICRAR M468, University of Western Australia, 35 Stirling Highway, Crawley, WA 6009}
\title{Automated detection of very Low Surface Brightness galaxies in the Virgo Cluster}

\date{Accepted XXX. Received YYY; in original form ZZZ}

\pubyear{2018}

\begin{document}
	\label{firstpage}
	\pagerange{\pageref{firstpage}--\pageref{lastpage}}
	\maketitle

\begin{abstract}
	
We report the automatic detection of a new sample of very low surface brightness (LSB) galaxies, likely members of the Virgo cluster. We introduce our new software, {\tt DeepScan}, that has been designed specifically to detect extended LSB features automatically using the DBSCAN algorithm. We demonstrate the technique by applying it over a 5 degree$^2$ portion of the Next-Generation Virgo Survey (NGVS) data to reveal 53 low surface brightness galaxies that are candidate cluster members based on their sizes and colours. 30 of these sources are new detections despite the region being searched specifically for LSB galaxies previously. Our final sample contains galaxies with $26.0\leq\langle \mu_{e}\rangle\leq28.5$ and $19\leq m_{g}\leq21$, making them some of the faintest known in Virgo. The majority of them have colours consistent with the red sequence, and have a mean stellar mass of $10^{6.3\pm0.5} M_{\odot}$ assuming cluster membership. After using {\tt ProFit} to fit S\'ersic profiles to our detections, none of the new sources have effective radii larger than 1.5 Kpc and do not meet the criteria for ultra-diffuse galaxy (UDG) classification, so we classify them as ultra-faint dwarfs. 

\end{abstract}


\section{Introduction}

The low surface brightness (LSB) universe is one that can be easily overlooked due to observational biases \citep{Disney1976}, yet remains an important test-bed for the enduring paradigm of the $\Lambda$CDM universe. It is theorised that a significant portion of the missing baryon budget \citep{Shull2012} may be hidden in diffuse sources such as intra-cluster light (ICL) \citep{Mihos2017}, tidal streams \citep{Cooper2010, Mowla2017} and LSB galaxies (LSBGs), an idea supported by the surprising abundance of ultra-diffuse galaxies (UDGs) originally detected in the Coma cluster \citep{VanDokkum2015, Koda2015}.  

\indent A great deal of effort has been devoted to searching for diffuse sources in other cluster environments, including the Virgo cluster \citep{Davies2015,Mihos2005,Mihos2015}; the Fornax cluster \citep{Kambas2000,Munoz2015} and others \citep[e.g.][]{Janssens2017,Roman2017}, as well as in the vicinities of massive galaxies \citep{Javanmardi2016,Muller2017}. LSB galaxies are also expected to form and exist in the field \citep{McGaugh1996,DiCintio2017,Amorisco2016} along with an ultra-diffuse intra-group baryonic component \citep{Driver2016}. 

\indent Of particular contemporary interest is the origin of the UDG. It is a matter of ongoing debate whether they can be described as ``failed L* galaxies'' \citep{VanDokkum2017}, tidally puffed-up dwarfs \citep{Collins2013} or the natural expectation of high spin dwarfs predicted from $\Lambda$CDM cosmology \citep{Amorisco2016}. Whatever the case, the issue has prompted several authors to search for LSB galaxies in a multitude of environments \citep{VanderBurg2016,VanderBurg2017, Roman2017}. One promising explanation has emerged indicating that UDGs may originate from dwarf galaxies that were quenched early on \citep{Beasley2016}, but statistical significance has been hampered by the inability to reliably detect very LSB objects other than by visual inspection \citep{Munoz2015,Venhola2017}. 

\indent Modern deep imaging surveys such as the Next Generation Virgo Survey (NGVS) \citep{Ferrarese2012}, Next Generation Fornax Survey (NGFS) \citep{Munoz2015}, KiDS \citep{DeJong2015}, and HSC-SPP \citep{Aihara2017} provide deep multi-wavelength data sets capable of probing the LSB Universe. Such data sets may offer a treasure trove of LSB objects, but it seems that current methodology has limited the ability for them to be fully exploited. While by-eye extraction of LSB sources currently has the very desirable advantage of minimal contamination from artefacts in the data, the sheer size of modern data means that only very small regions can be analysed this way. Upcoming surveys such as Euclid and LSST make this approach completely infeasible for the future.

\indent One of the main issues for LSB science is the quality of the data and its reduction. \cite{Slater2009} have made progress in removing artefacts caused by internal reflections of bright stars by modelling the point spread function (PSF) to high accuracy and out to wide radii. Further progress has been made in constructing very deep image stacks with special attention to preserving LSB features \citep{Blanton2011,Fliri2015,Mihos2017}. Despite these advances, many efforts to identify LSB features automatically have relied on methods that are sub-optimal for the extraction of low signal-to-noise ratio (SNR) sources. The most common of these is  {\tt SExtractor} \citep{Bertin1996a}, which at its core identifies objects as contiguous regions above the sky distribution at some confidence level. \cite{Davies2015} have shown that it tends to systematically miss or fragment large, diffuse objects even with settings optimised to the detection of LSB objects.

\indent Despite its popularity, there are alternatives to {\tt SExtractor}. Some examples include {\tt Clumpfind} \citep{Williams1994}, {\tt FellWalker} \citep{Berry2015} and others, but again these are not optimised to detect LSB objects and suffer from similar issues to {\tt SExtractor}. This has motivated several authors to create their own algorithms such as {\tt Oddity} \citep[and references within]{Butler-Yeoman2016}, {\tt MTObjects} \citep{Teeninga2016} and {\tt NoiseChisel} \citep{Akhlaghi2015}. An increasingly popular approach \citep[e.g.][]{Zheng2015} is to detect objects via a watershed segmentation algorithm, whereby detections can grow larger and larger until they reach a saddle point in intensity. This approach is also used by the  {\tt ProFound}\footnote{https://github.com/asgr/ProFound} software \citep{Robotham2018}.

\indent The above alternatives (with the exception of {\tt ProFound}, which is new) have not yet been widely used in the literature. Given the current abundance of research into the LSB Universe, we have been prompted to develop another detection package, called {\tt DeepScan}. This has been developed to meet the following criteria:

\begin{itemize}
	\item The software should supersede {\tt SExtractor} in its ability to detect extended LSB structure.
	\item Detection limits should be quantifiable so that the completeness of samples can be estimated.
	\item The algorithm should adapt to the various shapes of LSB sources e.g. LSBGs and ICL.
	\item The software should be intuitive and simple to use in a modern scripting language.
	\item The inputs and outputs of {\tt DeepScan} should be compatible with those of other software.
	\item The algorithm should be as efficient as possible and be written to take advantage of parallel  processing methods. 
\end{itemize}

\indent Our software implementation has been developed to run as efficiently as possible in order to cope with present and futuristic big-data challenges. The public availability of deep, wide-area survey data such as that of KiDS and VIKING \citep{Edge2013} (each covering thousands of square degrees with sub-arcsecond pixels) means that {\tt DeepScan} can be used immediately. Such survey data will make good testing grounds for the software, which we hope to apply to even larger surveys such as Euclid and LSST.

\indent This paper is intended to inform the reader of the methodology behind {\tt DeepScan}. We will endeavour to provide an up-to-date user's manual where the code is hosted publicly at {\tt https://github.com/danjampro/DeepScan} with a GPLv3 license. The paper is organised in the following way: In section \ref{section:DBSCAN}, we give an overview of the algorithm at the core of the detection method. In section \ref{section:deepscan} we give a brief overview of the {\tt DeepScan} software, with the full documentation made available online. We note that during the development of {\tt DeepScan}, \cite{Greco2017} have also developed a pipeline aiming to detect low surface brightness features using {\tt SExtractor}. While this work is quite similar in its objectives and methods, some of the differences between the two pieces of software are discussed in section \ref{section:sextractor}. In section \ref{section:NGVS} we give an example of its application to the publicly available NGVS data to reveal a sample of exceptionally faint galaxies. The sample retrieved from this analysis can provide a training-set for comparisons with other methods like {\tt ProFound}. Finally we discuss plans to increase the area of sky we have explored in this work. 

\indent While in this paper we primarily discuss the identification of LSB galaxies, the detection method employed by {\tt DeepScan} makes no assumptions about the underlying morphologies of its detections. It can therefore be used to detect other extended LSB sources such as tidal streams or intra-cluster light.

\section{DBSCAN in astronomy}
\label{section:DBSCAN}

DBSCAN \citep[Density-based spatial clustering of applications with noise,][]{Ester1996}, is a two-parameter algorithm that is designed to identify regions of high density within an $n$-dimensional data set. The algorithm has found  recent use in astronomy through the classification of eclipsing binaries \citep{Kochoska2017} and the morphological analysis of open clusters \citep{Bhattacharya2016}. Broadly speaking, our application of DBSCAN operates in a similar way to {\tt SExtractor} in that it builds detections by clustering together nearby pixels above some brightness threshold. In both cases, the detections are statistically unlikely to occur due to fluctuations in the background. 

\indent The fundamental difference between our use of DBSCAN and {\tt SExtractor} is that a DBSCAN detection is based on the density of pixels above a SNR threshold within its $\epsilon$ radius, whereas {\tt SExtractor} builds its detections by identifying contiguous regions of pixels with significant flux, possibly on a smoothed image. (We note that {\tt SExtractor} can amalgamate non-contiguous sources in its ``cleaning'' stage, which attempts to remove noise peaks that have been detected in the halos of brighter objects). 

\begin{figure}
  \includegraphics[width=\linewidth]{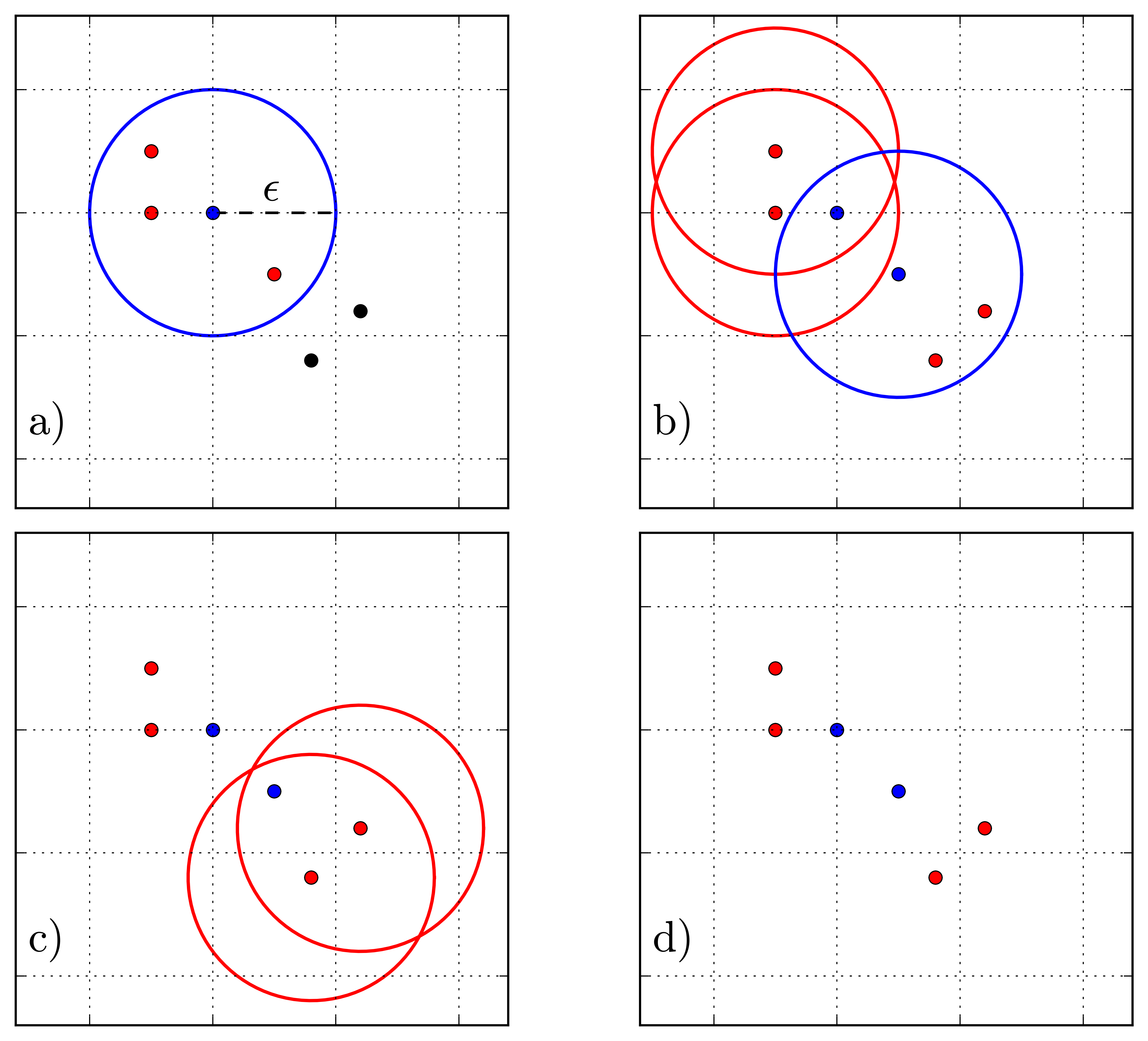}
  \centering
  \caption{A visualisation of the DSBSCAN algorithm with $\eta$=3. In a), the algorithm proceeds to identify a core point (blue). Then in b) it checks each of the secondary points (red) to see if they meet the criteria to be a core point - this is true of one of the secondary points, which is also shown in blue. The secondary points of the new core point are checked in c), and the final cluster, consisting of two core points, is shown in d).}
  \label{fig:dbscan}
\end{figure}

\indent The first parameter of DBSCAN is a clustering scale length ($\epsilon$) and the second ($\eta$) is the minimum number of data points required within an $\epsilon$ radius for a cluster to form. The algorithm iterates over every input data point. If the number of points $\varphi$ within a circular radius of length $\epsilon$ meets the condition  $\varphi\geq\eta$, the point is marked as a {\textit{core}} point; this is the basic building block of a cluster. Then, each point enclosed within $\epsilon$ (known as {\textit {secondary}} points) is checked to see if they also meet the condition to be core points, and if so then they are added to the same cluster. Thus, a cluster can contain more than one core point. This process repeats until there are no more core points to add to the cluster and it is complete. The algorithm then repeats the process to identify separate clusters within the dataset, if they exist. The clustering process is illustrated in figure \ref{fig:dbscan}.\newline
\indent Our approach is to use the spatial coordinates of pixels above a brightness threshold as inputs to DBSCAN, essentially identifying sources as over-densities of these pixels. This is analogous to how resolved LSB galaxies are detected through over-densities of their stars against the background. While we have not implemented usage of the upper detection threshold $I_{max}$ in {\tt DeepScan}, we include it in the modelling for completeness.

\indent The circular nature of the search aperture used in DBSCAN means that the algorithm has a ``resolution'' determined by $\epsilon$. This limitation does not rule out the detection of elongated structures if they are significant over scales similar to or larger than $\epsilon$. The algorithm therefore performs poorly in identifying objects significantly smaller than the detection circle and in separating sources closer together than 2$\epsilon$. While the former point is addressed by allowing the sensitivity of the algorithm to be set by the user, the latter could be remedied with a de-blending or segmentation routine. We have not implemented such a routine as instead we rely on the low spatial density of LSB objects and a high quality source mask to mitigate source confusion (see $\S$\ref{section:mask}). 

\indent The sensitivity of DBSCAN is set by $\eta$. To derive a value for $\eta$, a value of $\epsilon$ is assumed that remains a hyper-parameter of the algorithm (i.e. a parameter that is set by the user). $\eta$ is estimated with the assumption that the noise brightness distribution is a zero-mean Gaussian of standard deviation $\sigma_{n}$, i.e.:

\begin{equation}
P(I_{n})=P(I_{n},\mu=0,\sigma=\sigma_{n}) 
\label{eq:noise}
\end{equation}

\noindent for noise intensity $I_{n}$, mean $\mu$ and standard deviation $\sigma$. If a brightness threshold is applied with lower and upper boundaries $I_{min}$ and $I_{max}$ ($\mu_{min}$ and $\mu_{max}$ in magnitudes per square arcsecond) respectively, the pixel-to-pixel noise distribution can be used to predict the probability $P_{thresh}$ of a background pixel with a true brightness of $I_{b}$ lying within the threshold:

\begin{equation}
P_{thresh} = P(I_{min}\le I\le I_{max}) = \int_{I_{min}-I_{b}}^{I_{max}-I_{b}}P(I_{n}'
)dI'
\label{eq:prob_thresh1}
\end{equation}

\noindent Thus, an accurate model of the background is also assumed. As the amount of noise per pixel is modelled as an independent random variable, the binomial distribution can be used to calculate the number of pixels expected to lie within the brightness threshold within a circular region of radius $\epsilon$. In the hunt for LSB objects, $\epsilon$ should be large so that it encapsulates a high number of pixels. Therefore the binomial distribution can be approximately represented by another Gaussian,
\begin{equation}
P(\varphi)=P(\varphi, \mu_{\varphi}, \sigma_{\varphi}) 
\label{eq:prob_number}
\end{equation}
\noindent with 
\begin{equation}
\mu_{\varphi}=P_{thresh}N_{pix} 
\label{eq:binomial_mean}
\end{equation}
\begin{equation}
\sigma_{\varphi}=(P_{thresh}(1-P_{thresh})N_{pix})^{\frac{1}{2}}
\label{eq:binomial_std}
\end{equation}
\noindent where $N_{pix}$ is the number of pixels enclosed by a circle of radius $\epsilon$, and $\varphi$ is the number of those pixels within the threshold. Equation \ref{eq:prob_number} can be integrated between $\varphi'$ and $N_{pix}$ to find the probability $P_{0}$ of $\varphi'$ or more pixels in the circle lying within the threshold:

\begin{equation}
P_{0} = \frac{1}{2}\left[\erf\left(\frac{N_{pix}-\mu_{\varphi}}{\sigma_{\varphi}\sqrt{2}}\right) - \erf\left(\frac{\varphi'-\mu_{\varphi}}{\sigma_{\varphi}\sqrt{2}}\right)\right]
\label{eq:prob_eta}
\end{equation}

\noindent Setting $\varphi'$=$\eta$ and rearranging for $\eta$, we obtain

\begin{equation}
\eta = \sigma_{\varphi}\sqrt{2}\erf^{-1}\left[\erf\left(\frac{N_{pix}-\mu_{\varphi}}{\sigma_{\varphi}\sqrt{2}}\right)-2P_{0}\right] + \mu_{\varphi}
\label{eq:eta_prob}
\end{equation}

\noindent We can replace the hyper-parameter $\eta$ with a new parameter\footnote{Users of our software can still opt to specify $\eta$ manually.} $\kappa$ , defined as the number of standard deviations (equation \ref{eq:binomial_std}) above the expected number of points enclosed within $\epsilon$. We can therefore write the somewhat simpler expression,
\begin{equation}
\eta =  \mu_{\varphi} + \kappa \sigma_{\varphi}
\label{eq:kappa}
\end{equation}
\noindent where $\kappa$ is in one-to-one correspondence with the probability $P_{0}$. \newline
\indent Equation \ref{eq:eta_prob} describes the number of pixels lying within the brightness threshold, within a circle of radius $\epsilon$ embedded within pure Gaussian noise. It is useful as it expresses $\eta$ as a function of the probability of that many pixels occurring due to noise, a probability that can be set arbitrarily low by increasing $\kappa$. It also allows the prediction of what should be detected by the algorithm. For example, In the context of galaxy detection the detectable region on the central surface brightness (CSB), magnitude plane can be calculated. The brightness profiles of galaxies are often described by the S\' ersic profile \citep{Graham2005}, which can be expressed as:

\begin{equation}
I(r) = I_{0}\exp{\left(\frac{-r}{h}\right)^{\frac{1}{n}}} + I_{b}
\label{eq:Sersic}
\end{equation}

\noindent for radius $r$, CSB $I_{0}$ ($\mu_{0}$ in magnitudes per square arcsecond), scale size $h$ and S\' ersic index $n$. $I_{b}$ is the background brightness. In analogy to equation \ref{eq:prob_thresh1}, the probability of finding a pixel within the brightness threshold is:

\begin{equation}
P_{thresh}(I(r)) = \int_{\Delta I_{min}}^{\Delta I_{max}}P(I_{n}')dI'
\label{eq:prob_thresh2}
\end{equation}

\noindent with

\begin{equation}
\Delta I_{\substack{min\\ max}}=I_{\substack{min\\ max}}-I(r)
\label{eq:deltaI}
\end{equation}

\noindent Equation \ref{eq:prob_thresh2} can be integrated over a circular region to obtain 

\begin{equation}
\varphi_{s} = \pi\int_{0}^{\epsilon}{r\left[\erf\left(\frac{\Delta I_{max}}{\sigma_{n}\sqrt{2}}\right) - \erf\left(\frac{\Delta I_{min}}{\sigma_{n}\sqrt{2}}\right)\right]dr}
\label{eq:Sersic_minpts}
\end{equation}

\noindent The condition for the galaxy to be detected is then simply:

\begin{equation}
\varphi_{s} \ge \eta
\label{eq:detect}
\end{equation}

Numerical approximations to this condition are shown for a variety of values of $\epsilon$ in figure \ref{fig:limits_eps}. We show similar plots for $\kappa$ and the S\'ersic index $n$ in the appendix. 

\indent The effects of the PSF have not been modelled here. While the PSF is in general non-analytical and varies between datasets (and even in the same dataset), we probe the effects of a typical seeing PSF for ground based wide-area surveys in $\S$\ref{section:test} and have found the effect to be negligible for our target sources.

\begin{figure}
	\includegraphics[width=0.9\linewidth]{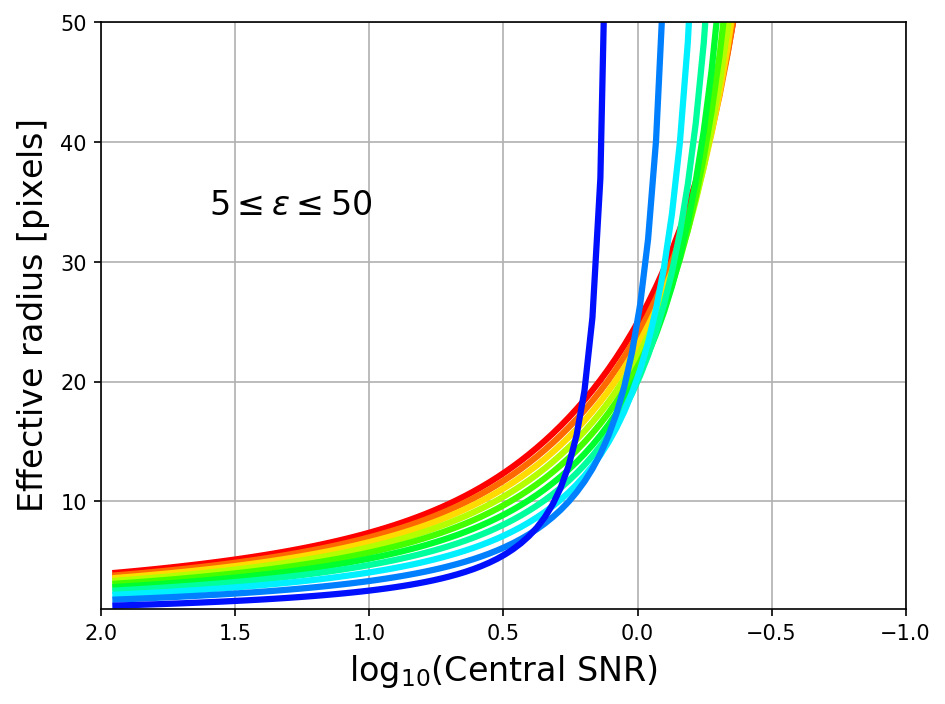}
	\centering
	\caption{Variation of the limits of detection with $\epsilon$, ranging between 5 (blue) and 50 pixels (red), with a colour gradient showing intermediate linearly-spaced values of $\epsilon$ increasing with redness. These results were obtained with $\kappa$=10, $I_{min}$=$\sigma_{n}$. The PSF and the effects of photon noise are not considered in this plot.}.
	\label{fig:limits_eps}
\end{figure}

\subsection{Testing DBSCAN}
\label{section:test}

In order to demonstrate the validity of the statistical modelling presented in section $\S$\ref{section:DBSCAN} we have performed artificial galaxy experiments, wherein sets of randomly generated circular S\' ersic profiles ($n$=1) were generated using {\tt ProFit}\footnote{https://github.com/ICRAR/ProFit} \citep{Robotham2017} and hidden in random noise of RMS=$\sigma_{n}$. $n$=1 was used because it is a fiducial value for dwarf galaxies \citep[e.g.][]{Koda2015} (but also see figure \ref{fig:limits_n}). We also include the effects of photon noise in the experiments and assume a gain of 1. A large grid of profiles was produced and embedded into random noise, using central surface brightnesses defined by their signal-to-noise (SNR) ratio (SNR logarithmically drawn between 0.1 and 100). The profiles have effective (half-light) radii between 1 and 15 pixels, where we have converted between S\' ersic quantities using the prescriptions of \cite{Graham2005}.

\indent Individual profiles were spaced by eight times the maximum effective radius of the sample, and were each truncated at 4 times this radius. This was to ensure that extended profiles could not contribute to their neighbour's detections. 

\begin{figure}
	\includegraphics[width=0.9\linewidth]{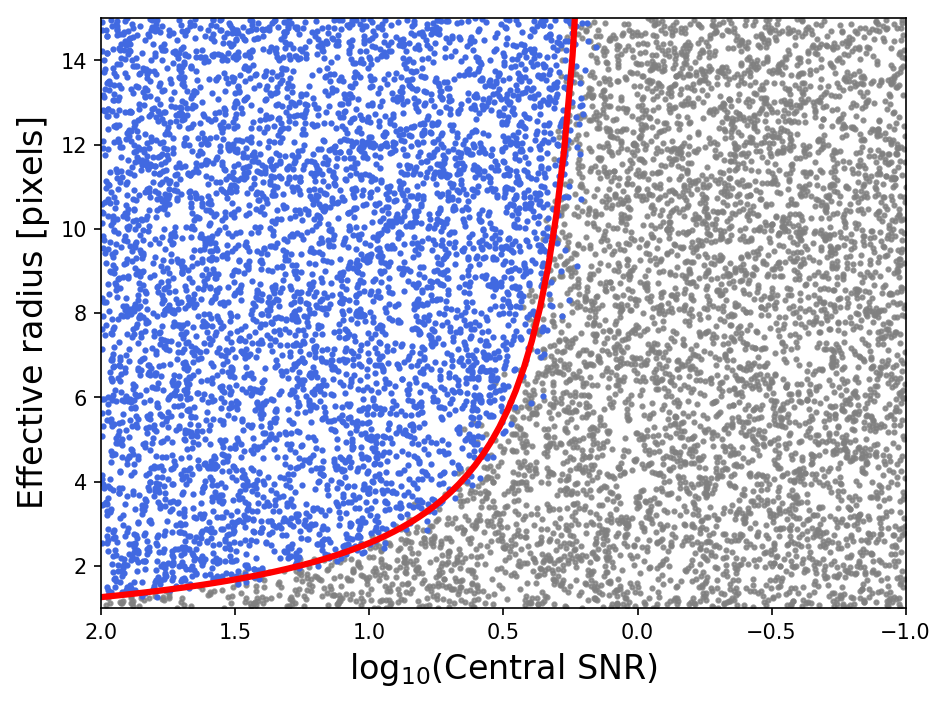}
	\centering
	\caption{Results of the synthetic profile experiment for $n$=1 circular S\' ersic profiles. Each point represents a synthetic source; blue ones have been detected and grey ones missed according to the criteria discussed in the text. The red line shows the boundary of detection predicted in equation \ref{eq:detect}. The sources presented here were not convolved with a PSF (see text).}
	\label{fig:synthetics}
\end{figure}

\indent Our new DBSCAN implementation (see $\S$\ref{section:deepscan_dbscan}) was then applied to the synthetic image, and any synthetic source that had an object located (by the mean coordinate of the core points) within twice its effective radius was regarded as detected. No two DBSCAN detections could be assigned to the same source and it was asserted that there were no DBSCAN detections that did not have matches. This check was to ensure that large portions of the image had not been detected as one. Results from matching the detections with their profiles are shown in figure \ref{fig:synthetics} for $\kappa$=10, $\epsilon$=5 pixels and a lower detection threshold of 1$\sigma_{n}$. Also on the plot is shown a numerical approximation (using the Nelder-Mead minimisation algorithm to the condition in equation \ref{eq:detect}). Importantly, the detection boundary predicted through the modelling is in good agreement with the observation of the synthetic data. 

\indent We have performed the same experiment after convolving the S\'ersic profiles with a mock Gaussian PSF with a full-width at half-maximum (FWHM) of 5 pixels (typical for 1$\arcsec$ seeing with a 0.2$\arcsec$ pixel size). The results were practically identical, even for the lower values of effective radii that we probed. This is easily explained from the fact that $\epsilon$ was also 5 pixels. Of course, as the PSF FWHM becomes larger than $\epsilon$, we would expect to see a retraction of the detection boundary. However, {\tt DeepScan} is intended for use on wide-area survey data (which typically has seeing and pixel scales of the order of what is probed here) and relatively large values of $\epsilon$, so this effect is not considered important.

\indent The preceding derivation of $\eta$ also does not consider the effects of correlated noise. This is noise that is ``clumpy'', produced during image stacking (interpolation, drizzling etc.) and from sources such as faint background galaxies that have not been accounted for in the sky modelling. In particular, this correlation will tend to make the uncertainty in the number of points contained within $\epsilon$ due to the sky distribution (equation \ref{eq:binomial_std}) an underestimate. The degree of noise correlation varies between datasets, making its effects difficult to quantify in general. For our current purpose of getting an estimate for $\eta$, the effects of underestimating $\sigma_{\varphi}$ can be accounted for by using higher values of $\kappa$, as can be seen in equation \ref{eq:kappa}. To estimate the degree to which $\kappa$ should change to accommodate the correlation, we generated independent random noise and applied a Gaussian filter with $\sigma$=1 pixel to create noise correlated over scales of two pixels. In such a set-up, Monte-Carlo trials suggest the standard deviation in equation \ref{eq:binomial_std} is underestimated by a factor of $\sim$2.5 and thus $\kappa$ would have to be multiplied by this factor to obtain equivalent behaviour to the uncorrelated case in terms of robustness against the detection of noise peaks (this comes at the cost of sensitivity). 

\indent Despite this, the assignment of $\kappa$ is likely to be done empirically rather than derived statistically because even with a source mask in place there will likely be unmasked sources contributing to a non-Gaussian background (see $\S$\ref{section:segmap}).

\indent There are similarities between DBSCAN and conventional data smoothing techniques because of the size of the search radius $\epsilon$. One major reason why very large smoothing kernels are not commonly used for detection is because nearby objects become confused with each other. Further, smoothing over bright, concentrated sources may produce detections that appear similar to LSB galaxies. By applying the source mask ($\S$\ref{section:mask}) \textit{before} the detection algorithm, this problem is alleviated and we can make use of the SNR obtained over larger areas without significant source confusion. DBSCAN is also more robust to the detection of small unmasked background objects because the input pixels are not flux-weighted; sources are forced to be significant over areas similar to the search area in order to be detected.\newline\newline\newline\newline

\section{The DeepScan software}
\label{section:deepscan}

{\tt DeepScan} is a Python package intended to identify regions of significant LSB light. The software uses a novel implementation of the DBSCAN algorithm that was created in order to operate much more efficiently than the standard. This efficiency is in-part due to many calls to integrated C code within {\tt numpy}\footnote{http://www.numpy.org} and {\tt scipy}\footnote{https://www.scipy.org}. One of the goals of {\tt DeepScan} has been to be compatible with other pieces of software, and as such there is a lot of flexibility as to what can be input to the software in terms of, for example, user-generated background maps or object masks. Equally, the outputs of {\tt DeepScan} such as segmentation maps and initial guesses on S\'ersic parameters can be easily transferred and used by different tools. If however the user does not have ready-made background maps etc., the basic usage of {\tt DeepScan}  is as follows: 

\begin{enumerate}
	\item Measurement of the sky distribution to produce sky and sky RMS maps (implicit source masking).
	\item Generation of a bright source mask (currently {\tt SExtractor} is used to create masks).
	\item Source detection on masked frames using DBSCAN.
	\item Automatic measurement of detections.
\end{enumerate}


There are two notable issues with this pipeline: 1) There is no source de-blending other than that of the source mask. The justification for this is that following appropriate source masking the spatial frequency of diffuse sources on the masked image is assumed to be low enough so that no de-blending is necessary; 2) Users are likely to want higher quality source measurements than the approximate ones provided with {\tt DeepScan}. We envisage these measurements to serve as inputs into robust profile fitting algorithms like {\tt ProFit}.
 
\subsection{Novel DBSCAN implementation}
\label{section:deepscan_dbscan}

An important feature of any detection algorithm is the runtime. Lower runtime helps users to fine-tune their parameters as well as enabling large data sets to be analysed over accessible (CPU) time-frames. Code optimisation usually proceeds by reducing the most significant time-consuming operation in the program; in the case of DBSCAN this is the region query, whereby the number of points within $\epsilon$ are counted for every data point. A simple implementation of DBSCAN may perform the region query by directly measuring the Euclidean distances from every input point to every other point and storing these distances in a symmetric distance matrix. This is inefficient in terms of memory as well as CPU time because every unique element of the matrix requires checking for every query. Indexing structures such as the R-tree \citep{Guttman1984} are optimised for spatial queries and allow for a significant speed-up. Many DBSCAN implementations use this method, including those in {\tt scikit-learn}\footnote{http://scikit-learn.org/stable/} and R\footnote{https://cran.r-project.org/web/packages/dbscan/dbscan.pdf}.

\indent However, our implementation is done in a much different manner that obtains equivalent results in notably shorter time-frames through a convolution approach. The basic procedure is the following:

\begin{enumerate}
	\item Create a binary image where pixels above the detection threshold are assigned the value of 1 and all others 0. We term the detection threshold as {\tt thresh}, which is quoted in units of the background RMS unless otherwise stated.
	\item Convolve the result with a top-hat filter of unit height and radius equal to $\epsilon$. This step essentially counts the number of thresholded pixels within an $\epsilon$ radius of every pixel.
	\item Threshold the resulting image at $\eta$ (derived from $\kappa$ and {\tt thresh}), creating a binary image with non-zero pixels being DBSCAN core points.
	\item Convolve the result with the same top-hat filter as in step 2. This connects regions corresponding to unique DBSCAN clusters. Set all non-zero pixels to 1.
	\item Run a contiguous pixel clustering algorithm over the result. This assigns unique integer labels \textgreater0 to each DBSCAN cluster. The result of this is known as {\tt segmap\_dilated}, and bounds the regions contained by all the DBSCAN core and secondary points.
	\item Perform a binary erosion on each object in  {\tt segmap\_dilated} to obtain the contiguous areas bounded by the core points. The result is simply called the {\tt segmap}. A segmentation map of only the core points ({\tt corepoints}) can also be retrieved.
\end{enumerate}

We note that an ``erosion" refers to contracting a source's segment with a kernel (in our case the top-hat filter of radius $\epsilon$); For each pixel making up a source's segment, all pixels that are contained within the kernel's footprint centred on that pixel are removed from the segment. A dilation is the opposite transformation. Hence the name ``{\tt segmap\_dilated}'' is appropriate because it can be obtained by performing a dilation on the {\tt segmap}. The only time an erosion is performed by {\tt DeepScan} is to create the {\tt segmap} from the {\tt segmap\_dilated}.

\indent The speed-up from the above approach compared to standard implementations stems from the fact that a) fast-Fourier transform (FFT) techniques can be used for the convolution steps and b) The need for the DBSCAN region query is removed and is replaced by a much more efficient contiguous-pixel clustering algorithm.

\indent We have tested {\tt DeepScan v1.0} against the {\tt scikit-learn} and {\tt R} DBSCAN implementations (versions 0.18.2 and 1.1-1 respectively) on one processor, and find that this implementation is faster than both. The tests were performed on a mid-2013 MacBook Pro (2.5 GHz Intel Core i5) with 8GB of RAM, running OSX 10.12.6. We also note that we find the R implementation to be significantly faster than that in {\tt scikit-learn}, but {\tt scikit-learn} gives the option to run DBSCAN in parallel whereas R does not. For example, averaging over five runs for a 500$\times$500 NGVS $g$-band cut-out, the times are {\tt DeepScan}: 0.3s, {\tt R}: 0.3s and {\tt scikit-learn}: 0.8s. Enlarging the image to 1000$\times$1000 pixels gives {\tt DeepScan}: 3.5s, {\tt R}: 6.0s and {\tt scikit-learn}: 17.6s. Scaling up once again to 4000$\times$4000 pixels, this time letting {\tt DeepScan} and {\tt scikit-learn} use four processors, the results are {\tt DeepScan}: 12.8s {\tt R}: 27.6s and {\tt scikit-learn}: 89.5s. We tested whether there was a difference between the output of {\tt DeepScan} compared to the other implementations and found that there was an exact match between the results.

\subsection{Sky measurement}

\label{section:sky}

\begin{figure}
	\includegraphics[width=0.9\linewidth]{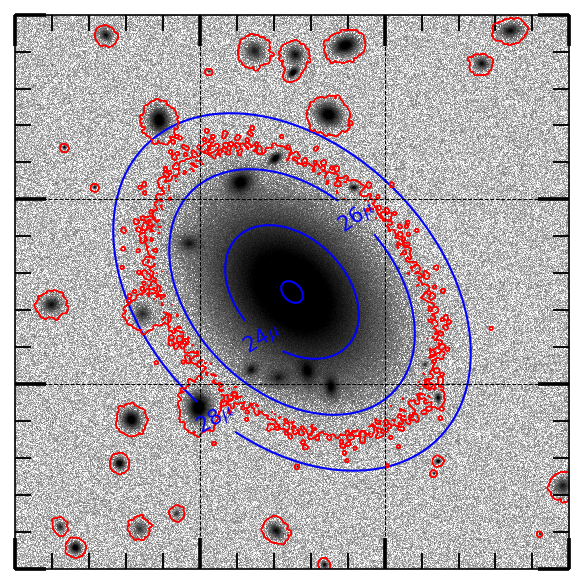}
	\centering
	\caption{This shows a synthetic 1500$\times$1500 pixel image, with an extended central source and some smaller ``background'' objects. Contours of the surface brightness profile are shown (blue) for the central profile, as well as a contour showing the masked region (red). Clearly the mask extends to very low surface brightness (the RMS of the image is 26.2$\mu$). The 500$\times$500 pixel meshes (black dashed lines) were median filtered over 3$\times$3 regions. The actual sky level of the image was set to zero ADU. Despite the presence of the dominant LSB object in the frame, the maximum value of the sky was measured to be 28.4$\mu$ - well below the RMS level. The RMS was recovered with an accuracy of over 99\%.}
	\label{fig:sky}
\end{figure}

{\tt DeepScan} can produce sky and sky RMS maps if required. To obtain an estimate of the sky we iteratively make measurements of the sky and the sky RMS, each iteration using DBSCAN with a low detection threshold (default {\tt thresh}=0.5) to identify sources (including LSB components) which are masked from the sky calculation in the following iteration. Using suitable values for $\epsilon$ (default of 5 pixels - similar to a typical PSF FWHM for wide field optical surveys) and $\kappa$ (default a value of 5 - low enough to encapsulate LSB components), this iterative masking reduces the bias incurred from unmasked LSB components each time. The iterations terminate when the sky level has converged to a user-specified tolerance. We have provided figure \ref{fig:sky} as an example of the sky-measurement algorithm, which was generated with default parameters. 

\indent The sky and sky RMS levels are estimated in meshes. The mesh size is a user-defined parameter. We maintain flexibility by allowing custom estimators for the measurements, but by default use the median for the sky and a lower-quantile estimate of the RMS (i.e. the level for which 15.9\% of the data is enclosed below the median). These are computed for each iteration, ignoring any masked pixels. The meshes are then median filtered over a customisable scale, before being interpolated over using a bi-cubic spline to the full image resolution. Meshes with too-few pixels are ignored in the interpolated over; by default at least 30\% of the mesh must be unmasked to count. Following this, DBSCAN is run, and any pixel identified within the {\tt segmap\_dilated} is masked. The algorithm then checks for convergence on a mesh-by-mesh basis; individual meshes that have converged are ignored for further iterations and their converged values are used in the interpolation. This process repeats with the updated source mask, either until all the meshes have converged or a maximum number iterations has been reached (the default is 6).

\indent We again emphasise that it is trivial to use sky and RMS maps generated externally from {\tt DeepScan}. We also note that custom masks can be used as an input to the source masking routine, which can be combined with the mask generated with DBSCAN or even treated as the final mask, in which the iterative mask generation is not applied.

\subsection{Masking bright sources}
\label{section:mask}

A crucial requirement of our detection method is a source mask. The mask must be created with the aim of eliminating all sources one does not wish to detect right out to their LSB halos. Our approach here has been to use {\tt SExtractor} to create the source mask. Measurements from the output catalogue such as the {\tt FLUX\_RADIUS} were found to do a poor job, underestimating the source sizes even with high values of {\tt PHOT\_FLUXFRAC} (the fraction of light contained within the flux radius). This prompted us to model each source with a S\'ersic profile and to size the ellipse according to some isophotal radius below the {\tt DeepScan} detection threshold.

\indent We estimated the S\'ersic index without performing any additional fitting from the ratio between the effective and Kron radii \citep{Graham2005}. This is useful as both of these measurements can be efficiently retrieved by {\tt SExtractor} with  the {\tt KRON\_RADIUS} and the {\tt FLUX\_RADIUS} keywords and {\tt PHOT\_FLUXFRAC}={\tt0.5}. Combining these measurements with the total magnitude (also measured using {\tt SExtractor}'s {\tt MAG\_AUTO}, by default with {\tt PHOT\_AUTOPARAMS} set to {\tt 2.5,3.5}), the profile is fully characterised. The source is then masked in an elliptical aperture (based on {\tt SExtractor}'s elliptical parameters) to the derived isophotal radius. We note that {\tt SExtractor} allows the possibility to perform S\'ersic fitting by requiring the {\tt SPHEROID\_SERSICN} or {\tt SPHEROID\_REFF\_IMAGE} columns in the output file. While we have not explored this in our current work, it is possible that this would improve the fits at the expense of some computation time.

A caveat of the above approach is that galaxies typically have non-elliptical LSB components and therefore may not be adequately masked. It is likely that {\tt ProFound} will eventually replace {\tt SExtractor} for the source masking as it offers the advantage of non-parametric object masks as well as more reliable estimates of parameters such as the half-light radius \citep{Robotham2018}.

\subsection{Source measurement}
\label{section:measurement}

The goal of {\tt DeepScan} is not to provide accurate profile fitting, but is rather to identify regions with significant LSB light. That said, we do provide a  basic function for 1-dimensional S\'ersic profile fitting in order to get initial estimates of parameters to input into robust 2D fitting programs such as {\tt ProFit}\footnote{\tt https://github.com/ICRAR/ProFit}. The basic requirements for the fit are the data and a segment corresponding to the source. 

\indent The initial task that is performed is the estimation of the centroid position and elliptical parameters (axis ratio and position angle). This is done using the same method of flux-weighted moments as in {\tt SExtractor} (see \cite{Bertin1996a} for detail), where the user has the choice to calculate the moments on either a masked or unmasked segmentation map. The user can choose which of the three segmentation maps provided ({\tt segmap}, {\tt segmap\_dilated} or {\tt corepoints}) to calculate these parameters.

\indent The next step is to measure the average (default median) flux within concentric elliptical annuli of fixed width centred on the source, based on the measurements from the previous step. The annuli iteratively increase their radius until a user-defined isophotal surface brightness is reached (often a relatively robust method of measuring large LSB galaxies) or a maximum radius has been reached. If the isophotal level is reached before sufficient steps have been performed then more steps will be taken in order to achieve a minimum number of data points (default is 5). 

\indent We then proceed to fit the profile using {\tt Scipy}'s {\tt curve\_fit} routine, which by default uses the Trust Region Reflective algorithm for parameter optimisation in the case of constrained problems (each S\'ersic parameter is constrained by default to have positive values). An initial parameter guess can be provided, but in its absence the parameters are estimated as following: The index $n$ is assumed as 1; The effective radius is assumed as the semi-major axis of the ellipse that bounds the segmentation map; the surface brightness at the effective radius is estimated by measuring the average surface brightness within the segmentation map, rescaling to effective surface brightness using the default value of $n$.

\indent We have not yet implemented methods to measure non-galaxy like objects. However, we suggest the  segmentation images outputted by {\tt DeepScan} can be used as inputs to non-parametric measurement tools such as that offered in {\tt ProFound} to provide estimates on parameters like total magnitude etc.

\section{DeepScan vs Source Extractor}
\label{section:sextractor}

\indent During our testing of {\tt SExtractor} we have found that it can perform fairly well in detecting LSB features provided specific input parameters are used. In this section we describe some observations about its usage and explain why {\tt DeepScan} may be preferred to detect specifically highly extended LSB objects. Our tests have consisted of us using a cut-out from the publicly available NGVS $g$-band data, which has a pixel size of 0.186$\arcsec$ and typical RMS of $\sim$26.9$g\mu$ with various combinations of {\tt SExtractor} settings. For the experiment, both {\tt SExtractor} and {\tt DeepScan} used the same background mesh of 50$\arcsec$ ($\sim$270 pixels) that was median filtered in 3$\times$3 meshes (Using the {\tt BACK\_SIZE} and {\tt BACK\_FILTERSIZE} {\tt SExtractor} keywords). The cut-out used had a size of 810$\times$810 pixels so that the background estimation was realistic. For this experiment, we have used {\tt SExtractor} to convolve the image with a Gaussian kernel of 5-pixels RMS, as in \cite{Greco2017}. 

\begin{figure*}
	\includegraphics[width=1\linewidth]{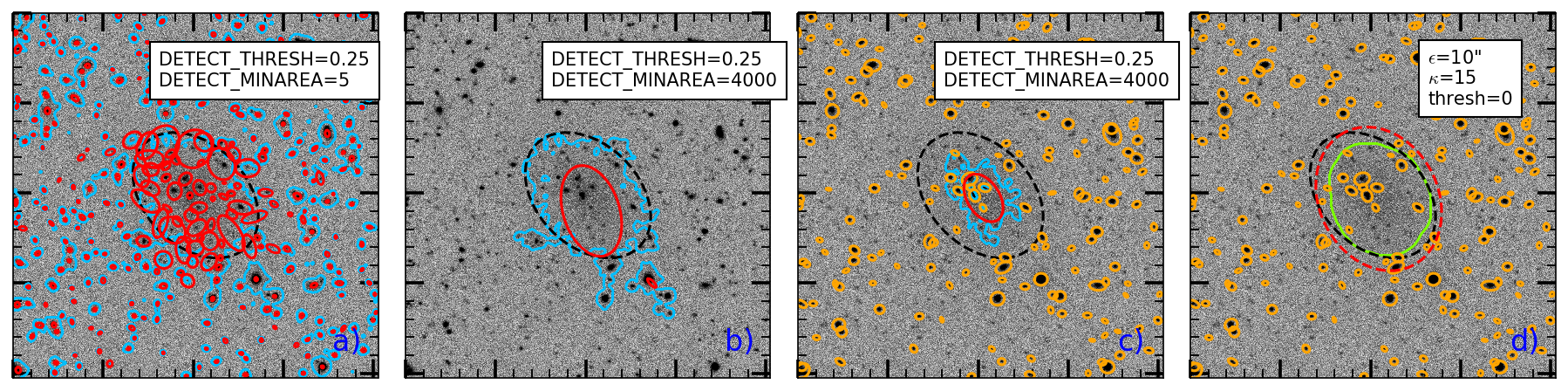}
	\centering
	\caption{This is a real 810$\times$810 pixel NGVS data cut-out with a synthetic LSB galaxy with a convolved central surface brightness of $\sim$26.9$g\mu$ and an exponential profile ($n$=1) of effective radius 30$\arcsec$ (roughly equivalent to the lower size limit of a UDG at Virgo). The black dotted ellipse represents this source out to one effective radius. In frames a) to c), the light blue contour traces objects on the {\tt SExtractor} {\tt SEGMENTATION} check plot. The red ellipses represent {\tt SExtractor} estimates of the effective radius for each detection, measured in the same way as in $\S$\ref{section:mask}. In figures c) and d), the orange ellipses bound the sources that are masked. In figure d), the green contour traces the {\tt DeepScan} {\tt segmap}, whereas the dashed red ellipse represents the effective radius as measured by {\tt DeepScan}. See text.}
	\label{fig:sex1}
\end{figure*}

\indent It is thought that the de-blending can routinely fragment large LSB features \citep{Davies2015, Greco2017} (aka ``shredding''). Indeed we have observed the de-blending fragmentation (figure \ref{fig:sex1}a), which was observed with  {\tt SExtractor}'s de-blending contrast parameter {\tt DEBLEND\_MINCONT}={\tt 0.005} (the default). A way around the problem is to deactivate the de-blending by setting {\tt DEBLEND\_MINCONT} to {\tt 1} - this value is used for the remainder of our tests. We have not experimented with different values of {\tt DEBLEND\_MINCONT} in this work. We also note that it is important for the {\tt CLEAN} parameter to be switched on (default) in order to reduce LSB source fragmentation, so it is activated in all our tests here. \newline\newline

\indent Figure \ref{fig:sex1}b) shows the result of  increasing the value of {\tt DETECT\_MINAREA} (the minimum number of pixels required for a detection to count) compared to a). A much better job is done of identifying the LSB source as a single object, but we note that the detection suffers two problems: the shape of the segment corresponding to the LSB source is significantly perturbed by background objects; and spurious detections still exist around groups of background objects despite very high values of {\tt DETECT\_MINAREA}. Activating the de-blending here exacerbates the situation and the LSB source is missed entirely, with its flux being solely attributed to some of the background objects rather than any central object.

\indent Herein lies a downside of using {\tt SExtractor} to detect LSB sources. The areas of the segmentation image corresponding to LSB objects detected on smoothed frames is made significantly unstable because of the presence of background objects that have either not been properly de-blended or have been erroneously assigned to the source in the cleaning stage. This is made clear by the morphology of the detection in figures \ref{fig:sex1}b) and c).  As this significantly effects the number of pixels an object contains, the usage of {\tt DETECT\_MINAREA} becomes an inherently unreliable tool to identify genuine LSB objects, yet is required to discriminate against background objects.
	
\indent The problem is partially alleviated by applying a source mask to the image before its input to {\tt SExtractor} as is also done by \cite{Greco2017} (note that there is no source mask handling within {\tt SExtractor}) in that now the only detection that appears in the output catalogue seems to be associated with the LSB object itself rather than brighter objects in its vicinity. However, the problems associated with the LSB segment still exist - the segment is irregular and contains several unmasked background objects that significantly perturb it's shape. It is also notable that the elliptical fit (determined by {\tt SExtractor}'s half-light  {\tt FLUX\_RADIUS} and elliptical parameters) does not do a reasonable job at measuring the object. This size underestimation can lead to the source being missed, as it is typical for authors to perform a cut on the minimum size of objects.

\indent In contrast, the {\tt DeepScan} detection (figure \ref{fig:sex1}d) is much smoother and does a better job of tracing the shape of the LSB structure. It is arguable that this is because we used a large value of $\epsilon$ (10$\arcsec$=50 pixels), but we note that {\tt DeepScan} is designed to work with such large kernels. The work of \cite{Greco2017} have also shown that using much larger kernels than 1$\arcsec$ is not feasible in {\tt SExtractor} because of blending with unmasked faint/background galaxies \citep{Koda2015, Sifon2018}, although this could be improved with a better masking strategy. We also note in passing that {\tt SExtractor} does not allow kernels larger than 31$\times$31 pixels, so it is impossible to use such a kernel from {\tt SExtractor}. A second point of consideration is that the core points of the {\tt DeepScan} detection define its shape, and these are identified as pixels with relatively high SNR on the original (i.e. non-smoothed) frame. Further, unmasked sources that cause the perturbations in the {\tt SExtractor} segmentation map have less of an effect on the shape of the {\tt DeepScan} detection because the pixels aren't flux weighted in the DBSCAN algorithm.

\indent To summarise, while it might be preferable to use {\tt DeepScan} to trace extended LSB light it is also possible to use {\tt SExtractor} to detect very low surface brightness objects, provided certain criteria are met:

\begin{enumerate}
	\item A ready-made source mask is provided.
	\item Large smoothing kernels are used (we note that the largest kernel size acceptable is 31$\times$31 pixels).
	\item Large values of {\tt DETECT\_MINAREA} are used.
	\item De-blending is deactivated.
	\item One treats parameters derived from the {\tt SEGMENTAION} check plot with some scepticism, particularly the {\tt FLUX\_RADIUS} as this seems to be systematically underestimated for large, diffuse objects. 
	\item Cleaning is on ({\tt CLEAN}={\tt Y}) to avoid spurious detections.
\end{enumerate}

\section{Application to the NGVS}
\label{section:NGVS}

To demonstrate the DBSCAN algorithm we applied it to a subset of the publicly available NGVS data that we acquired from the Canadian Astronomy Data Centre\footnote{http://www.cadc-ccda.hia-iha.nrc-cnrc.gc.ca/en/}. This data was taken with the square-degree MegaCam instrument on the Canada France Hawaii Telescope, and covers $\sim$100 square degrees of the Virgo Cluster in the $u,g,r,i,z$ bands. The NGVS was chosen because it offers deep imaging of the Virgo cluster at high resolution (0.186$\arcsec$ pixels), and was the same data used by \cite{Davies2015} with which we wish to compare. We use the $g$-band data as it has the best coverage, a low maximum seeing FWHM (1$\arcsec$) and an extended-source limit of 29$\mu$ \citep{Ferrarese2012}. The subset covers a five-degree$^{2}$ area projected radially eastwards from the centre of the Virgo cluster (i.e. M87). The subset is made from five overlapping frames, each covering an area of ~1 degree$^{2}$ with corresponding sizes of 21000$\times$20000 pixels. The frames are each 1.74Gb in size. This area overlaps with part of the region explored by \cite{Sabatini2005} so comparisons can also be made with their work. Objects detected in this region likely belong to sub-cluster A, the largest sub-cluster in Virgo \citep{Mei2007}. 

\indent Our general strategy is to use {\tt SExtractor} to identify sources for masking, before using {\tt DeepScan} to search the remaining area for LSB objects. The following processes were performed using four 2.5 GHz Intel Core i5 processors with 8Gb of RAM.  The overall pipeline we used is as follows:
 
\begin{enumerate}
	\item {\tt SExtractor} source masking
	\item Sky modelling
	\item Source detection
	\item Source selection
	\item Human validation
	\item S\'ersic fits with {\tt ProFit}
\end{enumerate}

\subsection{Source masking}

\begin{figure}
	\includegraphics[width=1\linewidth]{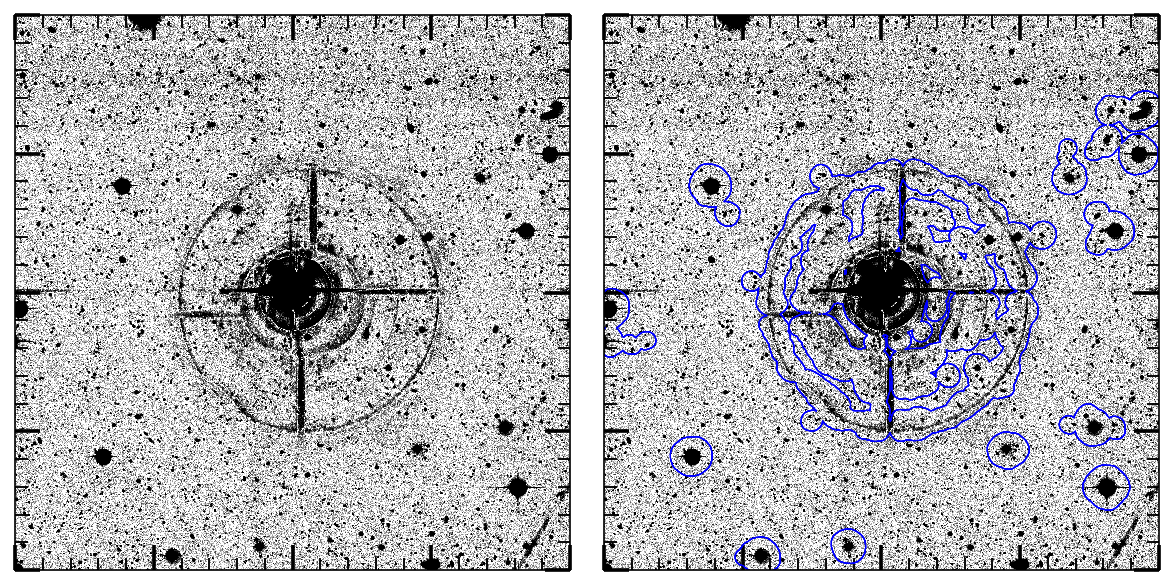}
	\centering
	\caption{Saturated star masking. The masked regions are contoured in blue against a smoothed cut-out of NGVS $g$-band data. The image size is 4000$\times$4000 pixels, or $\sim$12.5$\times$12.5 arcminutes. The large mask in the centre successfully masks out the LSB ring around the saturated star and we find no detections in its vicinity.}
	\label{fig:starmask}
\end{figure}

The first stage in the mask generation was to run DBSCAN over the raw data in order to identify saturated stars and their associated LSB halos. This is done because the {\tt SExtractor} masks generated for such objects were not sufficient to cover the sources. The parameters we used were $\epsilon$=10$\arcsec$ ($\sim$50 pixels), {\tt thresh}=0, $\kappa$=20. These are similar to those that we used for the actual LSB detection, but were modified based on trial and error masking of large saturated stars. Any detection that contained a saturated pixel within its {\tt segmap\_dilated} was masked within it. An example of the result of this saturated star masking is shown in figure \ref{fig:starmask}.

\indent The masked regions were set to zero and this data was used as the input to {\tt SExtractor}. For this we used a {\tt DETECT\_THRESH} of 6 (see $\S$\ref{section:segmap}) and convolved the image with the default filter (a 5$\times$5 pixel Gaussian filter of FWHM 2 pixels). This makes us sensitive to sources with surface brightnesses $\geq\sim$25$g\mu$ for the final DBSCAN run. We disabled de-blending in order to prevent the fragmentation of sources close to the detection threshold as this produced poor masks in their vicinities. We allowed {\tt SExtractor} to perform its own background and RMS estimates in small meshes of size 64$\times$64 pixels to better detect smaller sources against their local background. All other parameters were left to their defaults. The isophotal radii were then calculated as described in $\S$\ref{section:mask} for the 29$g\mu$ isophote.

\indent In a minority of cases, the initial {\tt SExtractor} mask did not cover the full source. This was particularly true for bright galaxies with extended LSB halos and bright unsaturated point sources where the approximate S\'ersic fits were inadequate. Requiring a fainter masking isophote did not solve the issue, so we were forced to enlarge the apertures by a factor of 1.5 for these sources (this was determined empirically).

\indent Each mask took approximately 30 minutes to generate. On average, 28 percent of each frame was masked excluding the border regions.

\subsection{Sky modelling}

We use the source mask as an input to {\tt DeepScan}'s sky modelling routine. We add to the map by masking sources detected by DBSCAN with the default paramters described in $\S$\ref{section:sky}; this allows sources to masked to well below the sky RMS level. We only do one iteration of the {\tt DeepScan} detection/masking process as the mask is already quite complete. The sky itself was measured in 500$\times$500 pixels and is median filtered over 3$\times$3 pixels. The large sky-mesh along with the median filtering increases our robustness against bias in the sky measurement due to unmasked LSB haloes. A second advantage is that 500 pixels ($\sim$90$\arcsec$) is large compared to the sources we are searching for so they should not be significantly subtracted with the sky.

\subsection{Source detection}

\label{section:segmap}

\begin{figure}
	\includegraphics[width=1\linewidth]{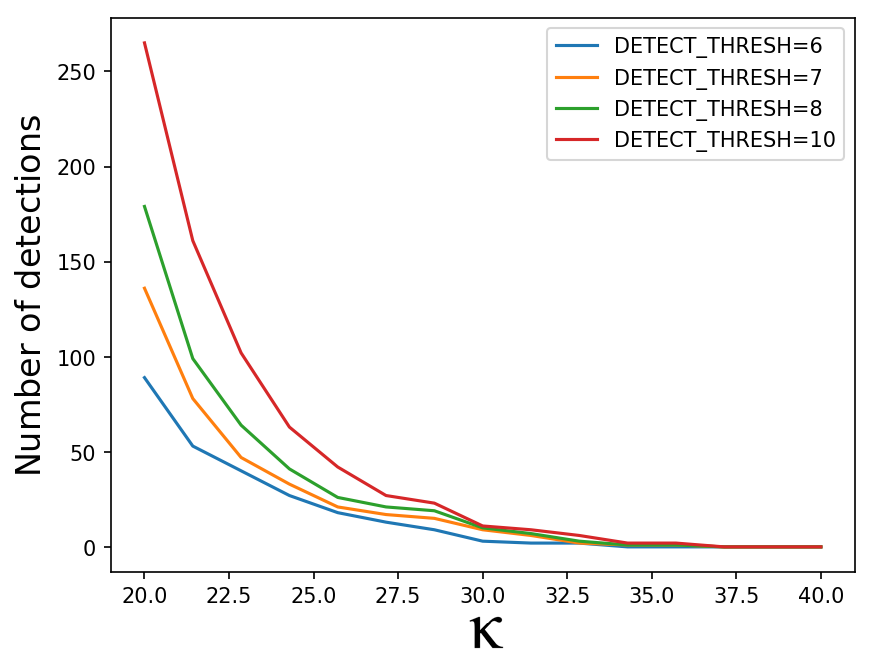}
	\centering
	\caption{{\tt DeepScan} parameter tuning. Here we show the effect of varying the {\tt SExtractor} detection threshold {\tt DETECT\_THRESH} and $\kappa$ parameter on the number of detections on a background NGVS frame. The number of sources detected begins to level off for $\kappa\geq30$ for each value of {\tt DETECT\_THRESH}.}
	\label{fig:kappas}
\end{figure}

The relevant parameters for the segmentation map generation using DBSCAN are the detection threshold, the search radius $\epsilon$ and $\kappa$. Our approach for setting the parameters was to perform empirical tests on a field image (this is an NGVS frame of an area of sky displaced from Virgo where we expect a low density of LSB objects). The detection threshold was set to 0.5 times the RMS ($\sim27.7g_{\mu}$) because lowering it much further made us sensitive to image artefacts such as background defects. An $\epsilon$ value of 10" ($\sim$50 pixels) was used as this is the smallest aperture that can critically sample an UDG at Virgo distance (UDGs have minimum size of 1.5Kpc \cite{VanDokkum2015} so at 16.5Mpc this is $\sim$20$\arcsec$). With these settings we can expect to detect UDGs with average SB within their effective radii of $\sim28g_{\mu}$. We ran the field image through the overall pipeline for several values of $\kappa$ and  {\tt SExtractor} detection thresholds (for the mask generation), with results shown in figure \ref{fig:kappas}. 

\indent From the figure, it is clear that the results converge for high values of $\kappa$. We select a value of $\kappa$=32.5 based on this plot by requiring less than 5 detections on the field image. With regards to the  {\tt SExtractor} detection threshold, we were interested in a value that was low enough in order to have a reasonably low number of contaminant objects while being high enough not to partially mask out LSB sources with shredded detections. We adopt a {\tt DETECT\_THREHSH} of 6 to lower the number of contaminant sources. Note that we did not probe lower values than 6 for the masking because this encroaches too far into the LSB regime, with surface brightnesses fainter than $\sim$25 magnitudes per square arcsecond.

\indent There is actually a significant difference in the background RMS level between the frames which makes us sensitive to different surface brightnesses from frame to frame. When we ran all the frames with the same settings as above, we found there was much more contamination of the output sample from spurious sources on some frames compared to others. We therefore normalised the threshold for each frame to the absolute surface brightness corresponding to that which was used on the field image, with settings as above. This is because lowering the threshold (in SNR) for frames with relatively high background RMS increases $\eta$ sufficiently to protect against the spurious detections.

\indent After setting the parameters, DBSCAN was run and took approximately 12 minutes per frame. In total, 67 objects were detected.

\subsection{Detection analysis}

\begin{figure*}
	\includegraphics[width=1\linewidth]{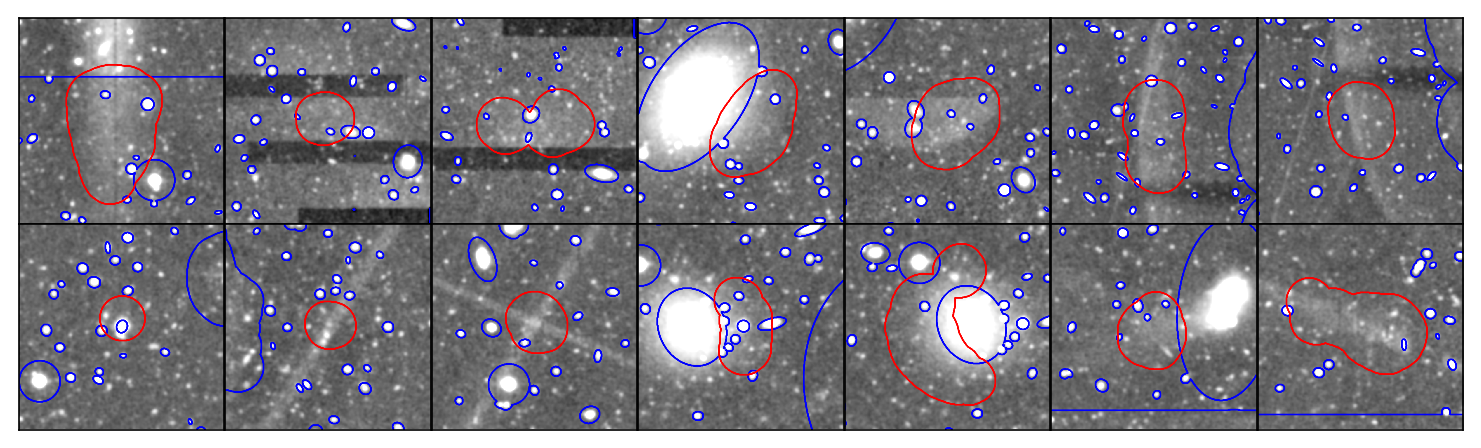}
	\centering
	\caption{Smoothed $g$-band cut-outs of the rejected detections. The maximum brightness of the dynamic range is approximately 25.5 magnitudes per square arcsecond. The red contour show the dilated segmentation maps produced with DBSCAN and the blue contour traces the mask.}
	\label{fig:rejects}
\end{figure*}

Each source was assessed visually in order to determine whether it was an astrophysical LSB source or miss-detection. We define miss-detections to encompass data artefacts (such as stellar diffraction rings and satellite trails) and the real LSB component of bright sources that have been inadequately masked. 14 of the raw detections were deemed to be miss-detections, leaving us with a sample of 53 objects. Of the rejected sources, 5 were associated with bright objects, 3 were unmasked stellar halos, 3 were satellite trails, 2 were caused by artefacts from the data stacking procedure and one was an extended LSB bloom caused by a bright source outside the FOV. The rejected sources are shown for clarity in figure \ref{fig:rejects}.

\indent The remaining sources were cross-matched with the VCC \citep{Binggeli1985}, LSBVCC \citep{Davies2015} and \cite{Sabatini2005} catalogues, using a search radius of 20$\arcsec$ (chosen so large to account for positional uncertainty in other surveys). 23 of the sources had matches, leaving a sample of 30 new LSB galaxies.

\indent We used {\tt DeepScan} to fit 1D S\'ersic profiles to the detections ignoring masked regions. We used the {\tt segmap\_dilated} to estimate elliptical parameters and the centroid positions. 500$\times$500 pixel cut-outs were obtained from the original data, the sky and RMS maps as well as the source mask and dilated segmentation map, centred on these centroids.

\indent The cut-outs were then used as inputs to the 2D Bayesian profile fitting package {\tt ProFit}, with initial parameter guesses given by the 1D fits. We follow the methodology suggested by the {\tt ProFit} team\footnote{http://rpubs.com/asgr/274695}, which consists of a three stage fitting process. First, a BFGS gradient decent fit is obtained. The results from this are then used as the initial parameters for a Laplace approximation using the method of Levenberg-Marquardt (LM). Finally, the results are used as the initial guesses for a more robust MCMC fit using the component-wise  hit-and-run metropolis (CHARM) algorithm, with 1000 iterations. In the fitting we used a simple Gaussian PSF of FWHM 1$\arcsec$. 

\indent The residuals for each fit were judged by-eye to ensure they were reasonable. In general they were, but for 8 of the sources we had to slightly modify the mask in order to get a good fit. Taking the standard deviation of the posterior distributions for each parameter result in uncertainty underestimates, likely because the high-quality of the initial parameter estimates and limited number of iterations mean only a narrow region of parameter space can be explored. To get a more realistic error, the upper and lower range of the posterior distributions were used as the parameter uncertainties. 

\indent In figure \ref{fig:cutouts} (appendix) we show the cut-outs for our final sample that contains new detections and those that had a match only in the catalogue of \cite{Sabatini2005}. We also provide a measurements table for these sources in the appendix. This is done because this catalogue does not cover the whole of Virgo in the same way that the VCC and LSBVCC do. These matches are of genuinely very low surface brightness and are obtained from a different dataset using a matched filter approach, making their re-detection a good coincidence test. We denote the names of galaxies in our final sample that had matches within the Sabatini catalogue with an asterisk after their name. On the figure we also show the elliptical annuli corresponding to the {\tt ProFit} fits out to the effective radius. Note that the galaxy VLSB23 does not have any measurements due to its highly unusual morphology, an odd over-density of point sources superimposed on a LSB fuzz, and may be worth investigating further.

\subsection{Results}

\begin{figure}
	\includegraphics[width=1\linewidth]{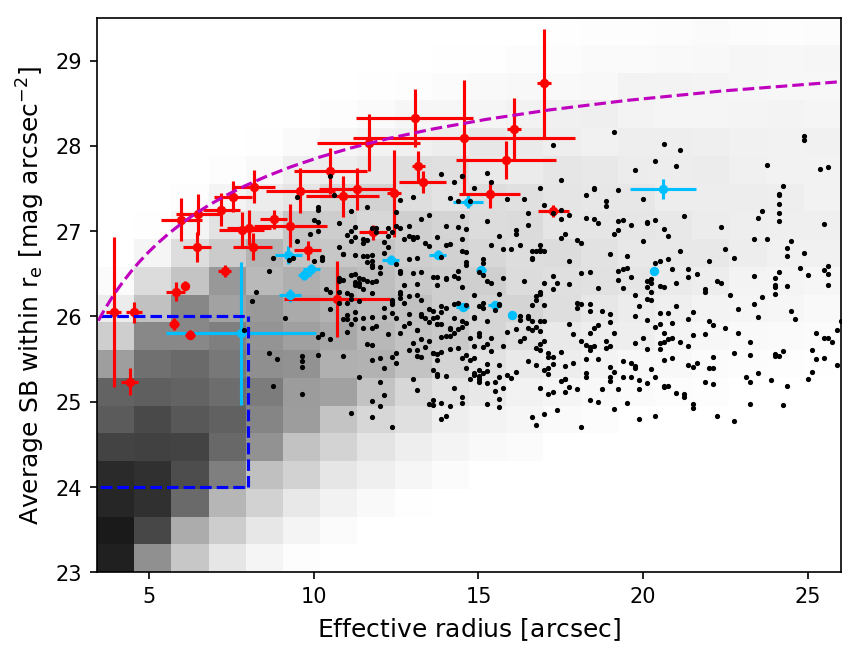}
	\centering
	\caption{Effective radius vs mean surface brightness within the effective radius for the new sample (red) and the extrapolated LSBVCC data (greyscale heatmap, see text). The light blue points represent the complementary sample. The blue box is the selection criteria used in \protect\cite{VanderBurg2017}, while the purple dashed line is the theoretical upper limit of detection given our  DBSCAN settings. The black dots represent the catalogue of \protect\cite{Yagi2016} projected at Virgo distance.}
	\label{fig:re_uae}
\end{figure}

\indent We plot the effective radius ($r_{e}$) vs the mean surface brightness within the effective radius ($\langle \mu_{e} \rangle$) based on our {\tt ProFit} models in figure \ref{fig:re_uae} for the final sample (includes matching \cite{Sabatini2005} sources). On the plot we also show the LSBVCC sample. However, the measurements presented in \cite{Davies2015} are in central surface brightness and exponential scale size units, and all assume a S\'ersic index of 1. To try and quantify the uncertainty this introduces on the $r_{e}$ - $\langle \mu_{e} \rangle$ plane, we take their initial results and calculate the relevant parameters using S\'ersic indices randomly generated based on our sample ($\langle n \rangle =1.0\pm 0.4$). We also plot the theoretical DBSCAN upper detection boundary assuming $n=1.4$ (i.e. 1$\sigma$ above the mean) which is consistent with our findings. 

\indent For context, we also show the selection criteria used by \cite{VanderBurg2017} in the figure, who used MegaCam imaging in the search for UDGs around groups and clusters. Further, we plot the sample of \cite{Yagi2016}, who obtained a catalogue of LSB galaxies in Coma with deep Subaru-R Suprime-Cam imaging, using their single S\'ersic {\tt GALFIT} \citep{Peng2002} fits. These results have been mapped to Virgo $g$-band data by assuming a Virgo distance of 16.5Mpc and a Coma distance of 99Mpc. The Subaru-R to $g$ conversion was done using a fiducial ($g-r$) value of 0.45 \citep{Roediger2017}.

\indent It is clear that the sample in this paper represents an extension of the parameter space explored by other surveys towards the very low surface brightness regime, with $\langle \mu_{e}\rangle > \sim 26.5$. It is perhaps surprising that no larger LSB objects were found and this may be in part due to the initial background subtraction performed on the public NGVS data, which is done over scales of 20$\arcsec$. We are hesitant to draw conclusions from this until we have a more complete sample, and completeness estimates, which we intend to acquire in a follow-up paper. 

\indent On the figure we also show the complementary sample, that is the sources that we detected but had matches in the VCC or LSBVCC. Two of these, VCC1331 and VCC1882, have measured effective radii larger than 20" and therefore may warrant reclassification to UDGs from their original classification of dwarf ellipticals. 

\begin{figure}
	\includegraphics[width=1\linewidth]{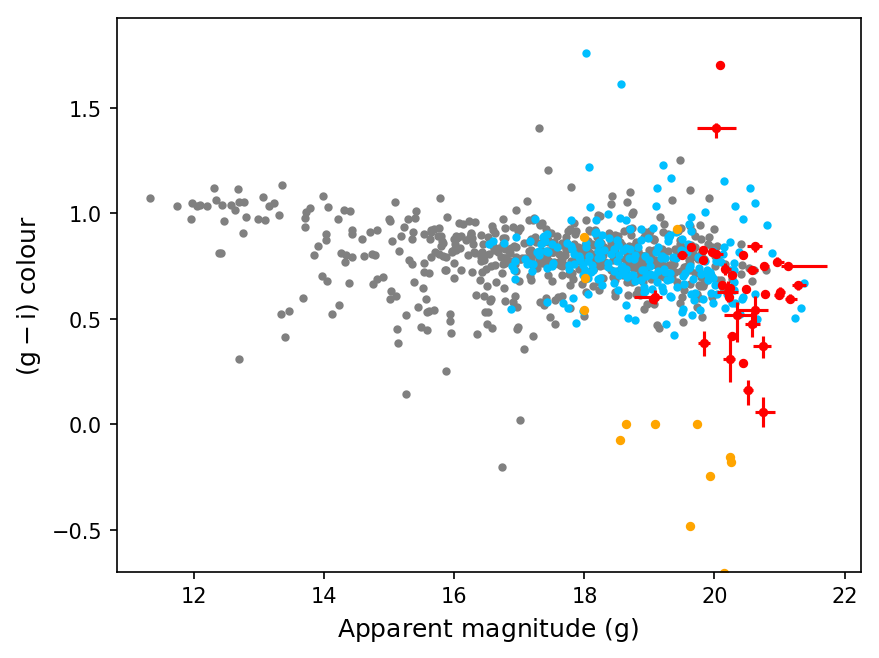}
	\centering
	\caption{Colour-magnitude diagram for galaxies in the VCC (grey), LSBVCC (blue) and the new sample (red). Many of the new sample are consistent with the Virgo red sequence, but there is a non-negligible sample occupying bluer colour space. The rejected detections are shown in orange.}
	\label{fig:col_mag}
\end{figure}

\indent We measured the ($g-i$) colours of the sample in elliptical apertures out to the effective radii measured in the $g$-band with {\tt ProFit}. For these measurements, we ignored the masked pixels, the results of which are shown in figure \ref{fig:col_mag}. For the $i$-band data we again used the publicly available NGVS data \citep{Ferrarese2012}. We also show measurements from the VCC and LSBVCC obtained by \cite{Keenan2017}. The general trend follows the Virgo red-sequence with fainter sources having redder colours and a flattening of the colour towards the faint end, as observed by \cite{Roediger2017}. Many of our final sample are consistent with this picture, but there are exceptions. Most noticeably, a collection of sources seems to depart the red sequence at the faint end, in favour of lower $(g-i)$ values. Note that this trend was still observed when recalculating the colours taking into account masked pixels. Two of the sources have unusually high values of ($g-i$). The source with the largest value (VLSB30) may have a biased colour due to its proximity to a star and the second, (VLSB19), seems to be associated with a large nucleated source. 

\indent Despite the red colours of the sources, they are generally better-detected in the NGVS $g$-band because fiducially the RMS level of $g$ is $\sim1.2$ magnitudes per square arcsecond fainter than that of $i$. 

\indent We briefly note that 10 of the 14 rejected detections have ($g-i$) colours below the minimum measured from our final sample, as can be seen in the figure. It may therefore be possible to increase the purity of the output automatically by applying a colour selection for future surveys. 

\begin{figure}
	\includegraphics[width=1\linewidth]{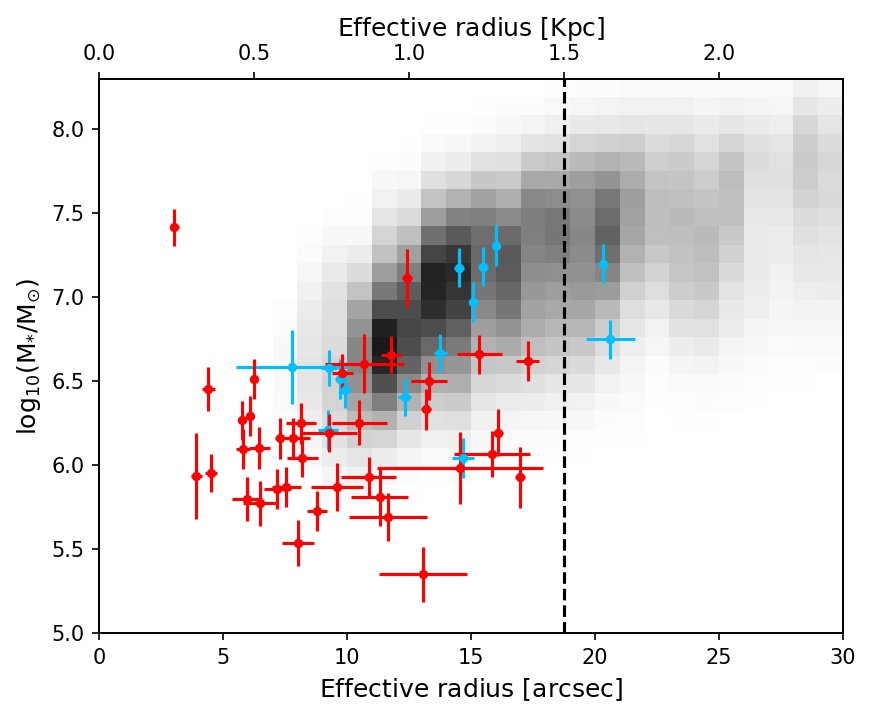}
	\centering
	\caption{Stellar masses vs effective radius for the final sample (red) and complementary sample (blue) assuming a Virgo distance of 16.5 Mpc. The heat map is the extrapolated data of \protect\cite{Yagi2016} projected at Virgo distance (see text). The black dotted line represents the size of the background estimation kernel in the original data reduction.}
	\label{fig:re_sm}
\end{figure}

\indent Figure \ref{fig:re_sm} shows the effective radii, both in units of arcseconds and Kpc (at the Virgo distance of 16.5$\pm$1.1Mpc \citep{Mei2007}) against the stellar mass calculated using the empirical relation derived by \cite{Taylor2011}. The galaxies have a mean (logged) stellar mass of $10^{6.3\pm0.5}M_{\odot}$, making them fairly less massive than the sample of UDGs presented in \cite{VanDokkum2015}, which have a median stellar mass of $6\times 10^{7}M_{\odot}$. Note that if the colours are measured without their source masks in place, the average stellar mass rises only slightly to $10^{6.4\pm1.0}M_{\odot}$. There is an outlier in the plot that corresponds to VLSB30 which is in proximity to a star. It is likely that the colour has been considerably effected by the star such that the stellar mass estimate may be erroneous.

\indent On the figure we have also plotted estimates of the stellar masses of the \cite{Yagi2016} sample projected at Virgo. Clearly there are several uncertainties in this procedure, but to attempt to get a representative picture we have randomly generated a set of data in which uncertain parameters have been perturbed within their errors, including the original error estimates from the {\tt GALFIT} models as well as uncertainties in distance and colour (we used $(g-i)=0.7\pm0.2$ based on figure \ref{fig:col_mag}). Our final sample seems to be both smaller in terms of size and also stellar content compared to their sample. It is interesting that some of our sources that matched with the VCC/LSBVCC agree well with the projected distribution from the Yagi sample, as it suggests that a re-inspection of the VCC/LSBVCC may result in the reclassification of some objects to UDGs.

\indent None of the final sample are larger than the 1.5 Kpc lower limit required for UDG classification; there is a notable dearth of large LSBs. Given their sizes and low stellar content, we classify them as ultra-faint dwarfs (UFDs). It could also be that the UDG population is already present in the catalogues of the VCC and LSBVCC but has not been explicitly identified as such; an idea supported by the fact that two of the galaxies in the complementary sample likely meet the UDG criteria. We note that the original NGVS background subtraction over 20" may have the effect of causing our measurements of the sizes of galaxies to be underestimates. 

\section{Discussion and Conclusions}

In this paper we have introduced a new software package that we have used to detect low surface brightness features in wide area survey data. The software is capable of measuring the background distribution and producing source masks, currently based on {\tt SExtractor} catalogues. The major novelty of {\tt DeepScan} is that it uses a highly efficient implementation of the DBSCAN algorithm to detect LSB features using much larger search radii than has been done before, allowing for the detection of extremely faint extended sources. 

\indent As with any detection process, there is a trade-off between completeness and the purity of the output sample. In {\tt DeepScan}, this is controlled by setting the DBSCAN input parameters: the clustering radius $\epsilon$, the confidence parameter $\kappa$ and the detection threshold {\tt thresh}. In general, larger values of $\epsilon$ allow for fainter objects to be detected, but using excessively large values may result in source confusion. $\kappa$ must be chosen high enough to protect against spurious detections of e.g. groups of background point sources, but setting it too high may result in unacceptably low completeness. 

\indent The purity of the output sample is dependent on the quality of the source mask. Creating a good mask is fairly difficult because of the problems involved in masking out LSB components associated with bright sources that do not adhere to elliptical profiles. Our current approach is to use {\tt SExtractor} to detect bright sources and mask out to an isophotal radius derived based on fitting a S\'ersic profile using outputs from the {\tt SExtractor} catalogue. This technique is successful in the majority of cases, but is not perfect. In the example application, we masked the LSB components associated with saturated regions using the {\tt segmap\_dilated} produced using DBSCAN. A disadvantage of this is possible source confusion, but is favourable because of its ability to mask large LSB features of arbitrary shape. One promising future approach to creating the source mask could be to use the dilation until convergence approach used by {\tt ProFound}, which can trace objects of arbitrary shape and thus provide non-parametric source masks.

\indent In the application to the NGVS data, the $\kappa$ value was chosen by measuring the number of objects detected on the frame as a function of $\kappa$ and choosing a value which had a low number of detections. Using such a high value of 32.5 means we have been limited in our capability to fully exploit DBSCAN because of the need to mitigate against contaminant sources in the output sample. Even with such a high value, we still reject 14 out of 67 detections, which consist mainly of satellite trails and unmasked regions associated with bright objects such as saturated stars. It is conceivable that some of these objects may be removed automatically using a colour analysis in future surveys on a larger scale.

\indent Of the remaining 53 sources, 30 do not have matches in either the VCC, LSBVCC or Sabatini catalogues. Keeping the Sabatini sources, we are left with a sample of 39. These measure to have parameter ranges of $26.0\leq\langle \mu_{e}\rangle\leq28.5g\mu$ and $19\leq m_{g}\leq21$ following fitting of S\'ersic profiles with {\tt ProFit}. Of this sample, none are large enough to be classified as UDGs and we classify them as UFDs (assuming cluster membership). Our current evidence for cluster membership is that they are reasonably consistent with the colours of other Virgo galaxies, and have angular sizes larger than the optimal selection criterion of $>$3" for Virgo galaxies. Assuming cluster membership, the galaxies have very low stellar masses, with an average of $10^{6.3\pm0.5}M_{\odot}$. 

\indent Comparing our final sample with those from other surveys, we find that we have probed a different region of parameter space, characterised by very low stellar mass estimates and surface brightness. We hypothesise that the dearth of larger detections stems from the initial background subtraction performed on the publicly available NGVS data. Following measurements of two galaxies VCC1331 and VCC1882, it is further hypothesised that some of the UDG population of Virgo may be contained within existing catalogues such as the VCC and LSBVCC.

\indent We have not made any efforts to estimate the completeness of our new sample. In future work we plan to perform a similar analysis on the whole of the NGVS. We aim to quantify the completeness by injecting synthetic sources into the data and measuring what we are able to recover in a similar way to that has been done by \cite{VanderBurg2017}; only then do we plan on drawing any astrophysical conclusions from our findings. The main conclusions from the experiment we have performed here is that we are able to detect new LSB features in areas that have specifically been searched for them before, which are some of the most diffuse detected in Virgo and reside in a different region in parameter space compared with those in the VCC and LSBVCC catalogues.

\section*{Acknowledgements} 
We are grateful for constructive comments provided by Dr. Michael Hilker \footnote{European Southern Observatory, Karl-Schwarzchild-Str. 2, D-85748 Garching, Germany}.\newline

\noindent We would also like to acknowledge the significant contribution the anonymous referee made to the quality of the paper. \newline

\noindent This work was supported by the Advanced Research Computing@Cardiff (ARCCA) Division, Cardiff University.

\bibliographystyle{mnras}
\bibliography{/Users/danjampro/Dropbox/phd/bibtex/awkward.bib,/Users/danjampro/Dropbox/phd/bibtex/library.bib}

\appendix
\section{Statistical limits of DBSCAN}

These figures are analagous to figure \ref{fig:limits_eps} but for the S\'ersic index $n$ and the DBSCAN detection threshold $\kappa$.

\begin{figure}
	\includegraphics[width=0.9\linewidth]{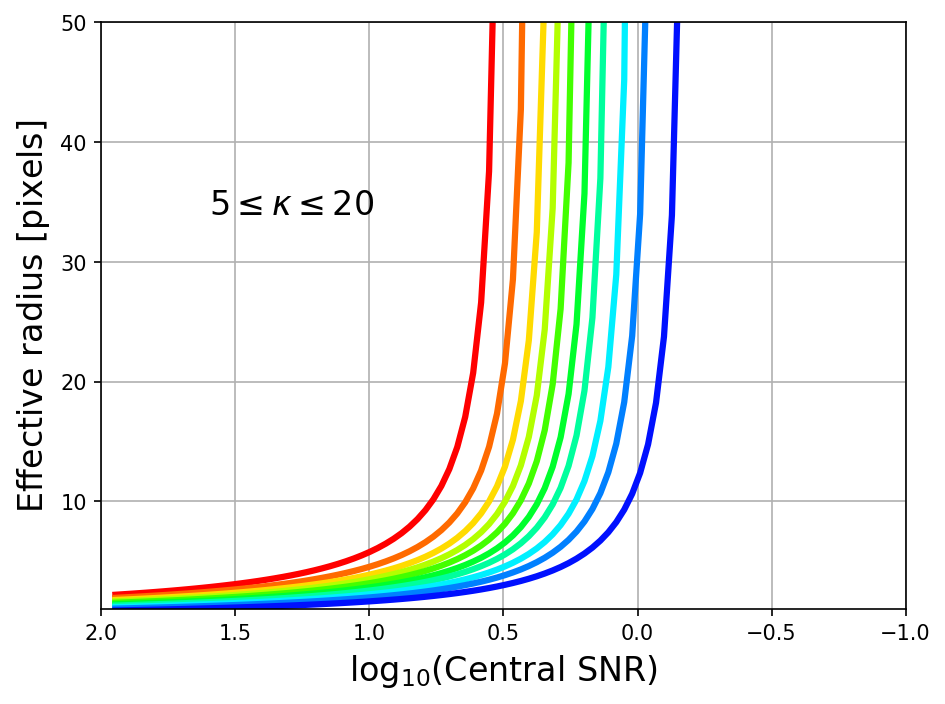}
	\centering
	\caption{Variation of the limits of detection with $\kappa$, ranging between 5 (blue) and 20 (red), with a colour gradient showing intermediate linearly-spaced values of $\kappa$ increasing with redness. These results were obtained with $\epsilon$=5 pixels, $I_min$=$\sigma_{n}$}.
	\label{fig:limits_kappa}
\end{figure}

\begin{figure}
	\includegraphics[width=0.9\linewidth]{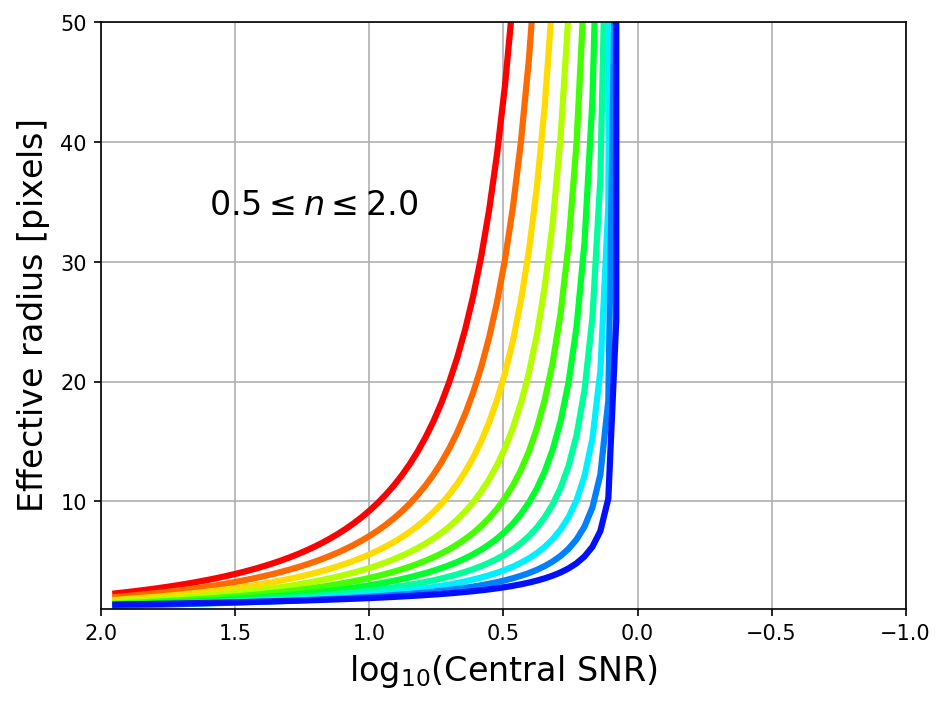}
	\centering
	\caption{Variation of the limits of detection with $n$, ranging between 0.5 (blue) and 2 (red), with a colour gradient showing intermediate linearly-spaced values of $n$ increasing with redness. These results were obtained with $\kappa$=10, $I_min$=$\sigma_{n}$, $\epsilon$=5 pixels}.
	\label{fig:limits_n}
\end{figure}

\section{Source cut-outs}

Smoothed data ($g$-band) cut-outs of our LSBG sample.

\begin{figure*}
	\includegraphics[]{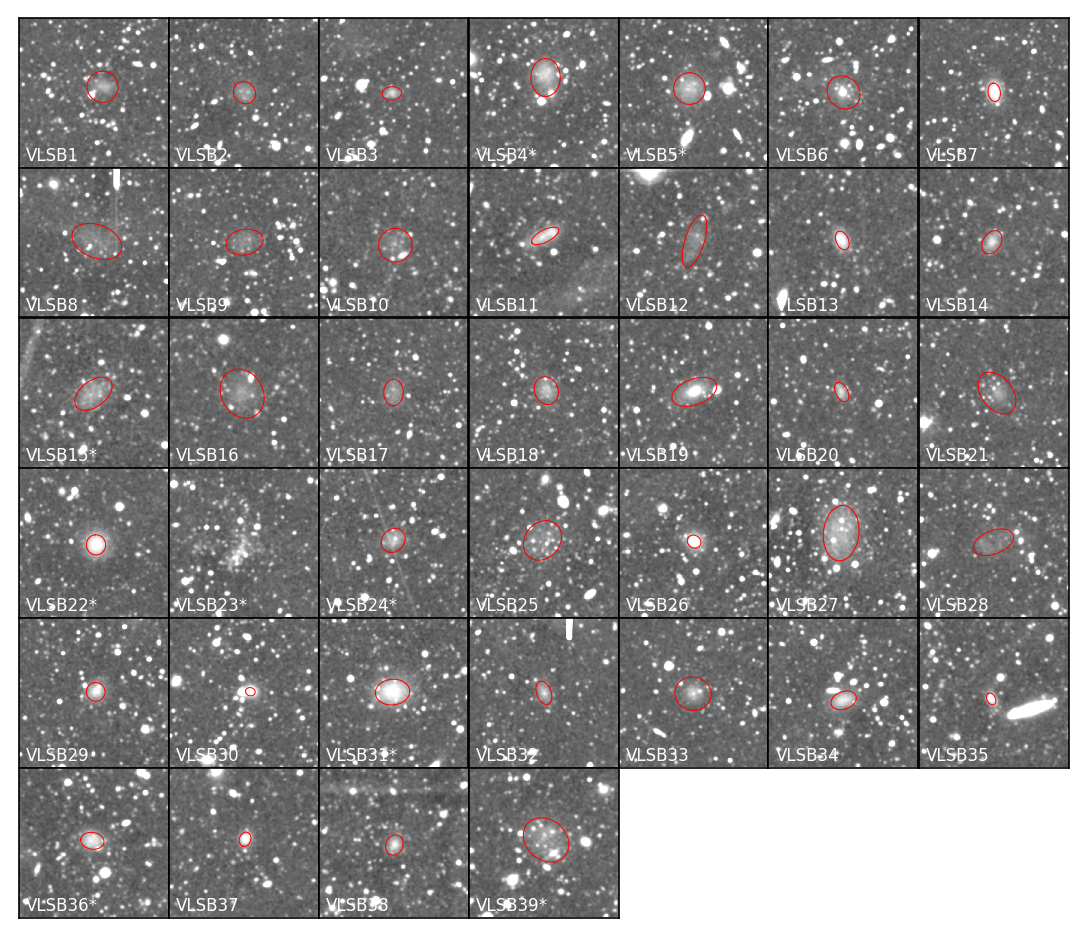}
	\centering
	\caption{Smoothed $g$-band cut-outs of our new detections. While some of the sample appear reasonably bright, the maximum brightness of the dynamic range is approximately 25.5 magnitudes per square arcsecond, which is close to the faint end limit of detection for the LSBVCC. The red ellipses show the {\tt ProFit} models out to the effective radii. Sources with an asterisk following their name are also present in the catalogue of \protect\cite{Sabatini2005}.}
	\label{fig:cutouts}
\end{figure*}

\section{Source measurements}

A table listing measured parameters and their errors for our LSBG sample. 

\centering
\begin{table*}
\begin{tabular}{| c | c | c | c | c | c | c | c| }
	
	Name & RA [deg] & Dec [deg] & mag ($g$) & $\langle \mu_{e} \rangle$ ($g$) & $r_{e}$ [arcsec] & $n$ & $(g-i)$ \\
	\hline
VLSB1 & 188.11080 & 11.62254 & 20.57 (0.13) & 27.47 (0.26) & 9.59 (1.04) & 1.27 (0.16) & 0.47 (0.06) \\
VLSB2 & 188.32165 & 11.62702 & 20.99 (0.08) & 27.25 (0.19) & 7.19 (0.55) & 0.78 (0.10) & 0.61 (0.02) \\
VLSB3 & 188.13531 & 11.70004 & 21.28 (0.13) & 27.13 (0.25) & 5.97 (0.62) & 1.26 (0.15) & 0.66 (0.02) \\
VLSB4* & 188.12484 & 11.83312 & 19.64 (0.04) & 26.99 (0.08) & 11.79 (0.42) & 0.94 (0.04) & 0.84 (0.01) \\
VLSB5* & 188.31528 & 11.86882 & 19.83 (0.05) & 26.77 (0.11) & 9.81 (0.41) & 0.90 (0.06) & 0.82 (0.01) \\
VLSB6 & 187.74268 & 11.97468 & 20.27 (0.09) & 27.41 (0.24) & 10.87 (1.11) & 1.99 (0.07) & 0.42 (0.02) \\
VLSB7 & 188.29238 & 12.08603 & 20.11 (0.03) & 25.91 (0.07) & 5.75 (0.15) & 0.92 (0.07) & 0.66 (0.01) \\
VLSB8 & 188.07065 & 12.12576 & 20.20 (0.08) & 28.20 (0.37) & 16.08 (0.20) & 0.60 (1.11) & 0.63 (0.05) \\
VLSB9 & 187.94944 & 12.30602 & 20.74 (0.15) & 28.04 (0.33) & 11.66 (1.57) & 0.74 (0.16) & 0.37 (0.06) \\
VLSB10 & 188.72908 & 11.60686 & 20.63 (0.13) & 27.70 (0.27) & 10.50 (1.11) & 0.74 (0.16) & 0.85 (0.03) \\
VLSB11 & 188.78022 & 11.65995 & 20.24 (0.06) & 27.06 (0.25) & 9.28 (1.10) & 0.62 (0.06) & 0.64 (0.02) \\
VLSB12 & 189.01353 & 11.71026 & 20.62 (0.11) & 28.74 (0.64) & 16.98 (0.21) & 0.58 (1.15) & 0.54 (0.08) \\
VLSB13 & 189.22741 & 11.84436 & 20.44 (0.02) & 26.36 (0.05) & 6.10 (0.12) & 0.83 (0.04) & 0.80 (0.01) \\
VLSB14 & 188.89157 & 11.91654 & 20.58 (0.09) & 27.02 (0.21) & 7.81 (0.68) & 1.21 (0.12) & 0.73 (0.01) \\
VLSB15* & 189.07497 & 11.95334 & 19.96 (0.06) & 27.58 (0.13) & 13.31 (0.71) & 0.73 (0.08) & 0.82 (0.01) \\
VLSB16 & 189.39290 & 12.10761 & 19.83 (0.09) & 27.83 (0.23) & 15.84 (1.52) & 0.97 (0.09) & 0.39 (0.06) \\
VLSB17 & 190.24144 & 11.60283 & 20.97 (0.09) & 27.52 (0.19) & 8.19 (0.63) & 0.67 (0.10) & 0.77 (0.02) \\
VLSB18 & 189.60967 & 11.65298 & 20.44 (0.05) & 27.14 (0.11) & 8.78 (0.41) & 1.01 (0.07) & 0.29 (0.02) \\
VLSB19 & 190.01774 & 11.83603 & 20.03 (0.22) & 27.44 (0.51) & 12.44 (0.20) & 1.33 (1.63) & 1.40 (0.05) \\
VLSB20 & 189.68450 & 11.94874 & 21.16 (0.11) & 27.20 (0.24) & 6.47 (0.67) & 1.13 (0.14) & 0.59 (0.02) \\
VLSB21 & 190.25336 & 12.03853 & 20.34 (0.28) & 28.10 (0.68) & 14.55 (3.37) & 1.32 (0.25) & 0.52 (0.13) \\
VLSB22* & 189.98940 & 12.11310 & 19.82 (0.02) & 25.78 (0.05) & 6.24 (0.14) & 0.78 (0.03) & 0.78 (0.00) \\
VLSB23* & 189.75714 & 12.20416 & -- & -- & -- & -- & -- \\
VLSB24* & 189.69605 & 12.23770 & 20.52 (0.09) & 27.04 (0.21) & 8.02 (0.66) & 1.02 (0.15) & 0.16 (0.07) \\
VLSB25 & 190.25519 & 12.28037 & 20.17 (0.04) & 27.76 (0.18) & 13.18 (0.19) & 0.53 (1.08) & 0.73 (0.03) \\
VLSB26 & 191.04250 & 11.57866 & 20.02 (0.10) & 25.23 (0.16) & 4.40 (0.27) & 1.88 (0.34) & 0.81 (0.02) \\
VLSB27 & 190.82000 & 11.62940 & 19.05 (0.03) & 27.24 (0.07) & 17.28 (0.46) & 0.63 (0.03) & 0.59 (0.01) \\
VLSB28 & 191.27329 & 11.66796 & 20.75 (0.18) & 28.32 (0.34) & 13.06 (1.79) & 0.34 (0.07) & 0.06 (0.07) \\
VLSB29 & 190.69944 & 11.69863 & 20.48 (0.05) & 26.29 (0.12) & 5.80 (0.28) & 1.15 (0.08) & 0.64 (0.02) \\
VLSB30 & 191.10218 & 11.94348 & 20.09 (0.03) & 24.48 (0.14) & 3.03 (0.18) & 2.50 (0.00) & 1.70 (0.00) \\
VLSB31* & 191.31792 & 12.24813 & 19.08 (0.32) & 26.20 (0.44) & 10.70 (1.60) & 1.44 (0.79) & 0.60 (0.03) \\
VLSB32 & 190.89820 & 12.32410 & 21.02 (0.07) & 27.40 (0.18) & 7.55 (0.58) & 1.03 (0.14) & 0.63 (0.02) \\
VLSB33 & 191.07622 & 12.33591 & 20.23 (0.10) & 27.49 (0.24) & 11.31 (1.15) & 1.86 (0.16) & 0.31 (0.11) \\
VLSB34 & 190.55282 & 12.37594 & 20.27 (0.04) & 26.82 (0.16) & 8.15 (0.59) & 0.92 (0.06) & 0.71 (0.01) \\
VLSB35 & 192.03936 & 11.59721 & 21.14 (0.29) & 26.05 (0.88) & 3.92 (0.23) & 0.89 (1.19) & 0.75 (0.01) \\
VLSB36* & 191.69418 & 11.65545 & 20.22 (0.03) & 26.53 (0.07) & 7.29 (0.21) & 0.75 (0.05) & 0.61 (0.01) \\
VLSB37 & 192.28541 & 11.91996 & 20.78 (0.06) & 26.05 (0.12) & 4.53 (0.25) & 1.14 (0.12) & 0.62 (0.01) \\
VLSB38 & 191.70495 & 11.94819 & 20.76 (0.08) & 26.81 (0.18) & 6.46 (0.43) & 1.40 (0.11) & 0.75 (0.01) \\
VLSB39* & 191.70132 & 12.19595 & 19.51 (0.08) & 27.43 (0.16) & 15.34 (0.91) & 0.71 (0.09) & 0.80 (0.01) \\
	
\end{tabular}
\end{table*}

\end{document}